\documentclass[fleqn,usenatbib]{mnras}

\usepackage{epsfig}
\usepackage{times}
\usepackage[british]{babel}             
\usepackage[slantedGreek]{newtxmath}    
\usepackage{threeparttable}

\usepackage{amsmath}
\usepackage{newtxtext,newtxmath}

\let\la=\lesssim 
\usepackage[T1]{fontenc}                
\usepackage{graphicx}                   

\usepackage{flexisym}
\usepackage{subfig}
\usepackage{hyperref}

\newif\ifAMStwofonts

\newcommand{\target}{Swift\,J1858.6--0814}

\newcommand{\ergs}{\,erg\,s$^{-1}$\,cm$^{-2}$}
\newcommand{\erg}{\,erg\,s$^{-1}$}
\newcommand{\Msun}{\,$\rm M_{\sun}$}
\newcommand{\Rsun}{\,$\rm R_{\sun}$}

\newcommand{\ucamS}{$u',g',i'$}
\newcommand{\ucam}{$u_s,g_s,i_s$}
\newcommand{\hcam}{$u_s,g_s,r_s,i_s,z_s$}
\usepackage{appendix}
\usepackage{footnote}

\makesavenoteenv{tabular}
\makesavenoteenv{table}

\title[An optical/X-ray timing study of Swift\,J1858.6--0814]
{A rapid optical and X-ray timing study of the neutron star X-ray binary Swift\,J1858.6--0814}

\author[T. Shahbaz et al.]{T. Shahbaz,$^{1,2}$\thanks{E-mail: tsh@iac.es}
	J. A. Paice,$^{3,4,5}$
	K. M. Rajwade,$^{5,6}$
	A. Veledina,$^{7,8}$
	P. Gandhi,$^{3}$
	V. S. Dhillon,$^{1,2,9}$\newauthor
	T. R. Marsh,$^{10}$
	S. Littlefair,$^{9}$
	M. R. Kennedy,$^{5,11}$
	R. P. Breton$^{5}$ and
	C. J. Clark.$^{5,12,13}$
	\\
  $^1$Instituto de Astrof\'\i{}sica de Canarias (IAC), E-38205 La Laguna,  Tenerife, Spain \\
  $^2$Departamento de  Astrof\'\i{}sica, Universidad de La Laguna (ULL),  E-38206 La Laguna, Tenerife, Spain \\
  $^{3}$Department of Physics and Astronomy, University of Southampton, Highfield, Southampton, SO17 1BJ, UK \\
  $^{4}$Inter-University Centre for Astronomy and Astrophysics, Pune, Maharashtra 411007, India \\
$^{5}$Department of Physics and Astronomy, The University of Manchester, Oxford Road, Manchester, M13 9PL, UK \\
$^{6}$ASTRON, the Netherlands Institute for Radio Astronomy, Oude Hoogeveensedijk 4, 7991 PD Dwingeloo, The Netherlands\\
$^{7}$Department of Physics and Astronomy, FI-20014 University of Turku, Finland\\
$^{8}$Nordita, KTH Royal Institute of Technology and Stockholm University, Roslagstullsbacken 23, SE-10691 Stockholm, Sweden\\
$^{9}$Department of Physics and Astronomy, University of Sheffield, Sheffield, S3 7RH, UK\\
$^{10}$Astronomy and Astrophysics Group, Department of Physics, University of Warwick, Gibbet Hill Road, Coventry, CV4 7AL, UK \\ 
$^{11}$Department of Physics and Astronomy, University College Cork, College Rd, Cork Cork T12 K8AF, Ireland\\
$^{12}$Max-Planck-Institut f\"ur Gravitationsphysik (Albert-Einstein-Institut), Callinstra{\ss}e 38, 30167 Hannover, Germany \\
$^{13}$Leibniz Universit\"at Hannover, 30167 Hannover, Germany
}

\date{Accepted XXX. Received YYY; in original form ZZZ}

\pubyear{2021}

\begin{document}
\label{firstpage}
\pagerange{\pageref{firstpage}--\pageref{lastpage}}
\maketitle


\begin{abstract} 
\noindent
We present a rapid timing analysis of optical (HiPERCAM and ULTRACAM) and X-ray (NICER) observations of the X-ray transient \target\ during 2018 and 2019. The optical light curves show relatively slow, large amplitude ($\sim$1\,mags in $g_s$) `blue' flares (i.e. stronger at shorter wavelengths) on time-scales of $\sim$minutes  as well as fast, small amplitude ($\sim$0.1\,mag in $g_s$) `red' flares (i.e. stronger at longer wavelengths) on time-scales of $\sim$seconds. The `blue' and `red' flares are consistent with X-ray reprocessing and optically thin synchrotron  emission, respectively, similar to what is observed in other X-ray binaries. The simultaneous optical versus soft- and hard-band X-ray light curves show time- and energy dependent correlations. 
The 2019 March 4 and parts of the June data show a nearly symmetric positive cross correlations (CCFs) at positive lags consistent with simple X-ray disc reprocessing. The soft- and hard-band CCFs  are similar and can be reproduced if disc reprocessing dominates in the optical and one component (disc or synchrotron Comptonization) dominates both the soft and hard X-rays. A part of the 2019 June data shows a very different CCFs. The observed positive correlation at negative lag in the soft-band can be reproduced if the optical synchrotron emission is correlated with the hot flow X-ray emission. 
The observed timing properties are in qualitative agreement with the hybrid inner hot accretion flow model, where the relative role of the different X-ray and optical components that vary during the course of the outburst, as well as on shorter time-scales, govern the shape of the optical/X-ray CCFs.
\end{abstract}


\begin{table*}
\centering
\caption{Log of ULTRACAM, HiPERCAM \& NICER observations for \target.}
\centering
\begin{tabular}{lcccccl}
\hline
UT date    & UT Start & UT End & Instrument & Filters      & Cadence$^1$     & Comments\\
\hline 
2018/11/14 & 19:24:22 & 19:50:15 & HiPERCAM & \hcam         & 46.6\,ms        & \\
2018/11/09 & 00:42:59 & 01:18:33 & ULTRACAM & \ucamS        & 0.93 (4.63)\,s  &  \\
2019/03/01 & 09:14:15 & 09:49:19 & ULTRACAM & \ucam         & 1.00 (3.01)\,s    &  \\
2019/03/02 & 09:07:00 & 09:39:20 & NICER    &  0.2--12\,keV & 40\,ns          & ObsId 2200400101	 \\
2019/03/02 & 09:05:22 & 09:45:14 & ULTRACAM & \ucam         & 0.28\,s (0.29)\,s        & Simultaneous with NICER \\
2019/03/04 & 08:54:22 & 09:48:59 & ULTRACAM & \ucam         & 0.50\,s (4.01)\,s         & Simultaneous with NICER \\
2019/03/04 & 09:05:25 &	09:36:20 & NICER    &  0.2--12\,keV & 40\,ns          & ObsId 2200400103	 \\
2019/03/05 & 09:06:26 & 09:26:49 & ULTRACAM & \ucam         & 0.58\,s (1.17)\,s       & \\
2019/05/09 & 08:13:38 & 10:23:17 & ULTRACAM & \ucam         & 0.25\,s (1.26)\,s   &   \\ 
2019/06/07 & 01:52:25 & 02:39:45 & HiPERCAM & \hcam         & 47.9\,ms        & Simultaneous with NICER \\
2019/06/07 & 01:53:20 &	02:33:43 & NICER    &  0.2--12\,keV & 40\,ns          & ObsId 2541030101 \\	
2019/06/07 & 03:23:09 & 04:17:27 & HiPERCAM & \hcam         & 47.9\,ms        & \\ 
\hline 
\end{tabular} 
\label{table:log}
\begin{tablenotes}
  \item $^1$Numbers in brackets is the $u_s$-band cadence, if different from the other wave-bands.
\end{tablenotes}
\end{table*}

\begin{figure*}
\includegraphics[width=1.0\linewidth,angle=0]{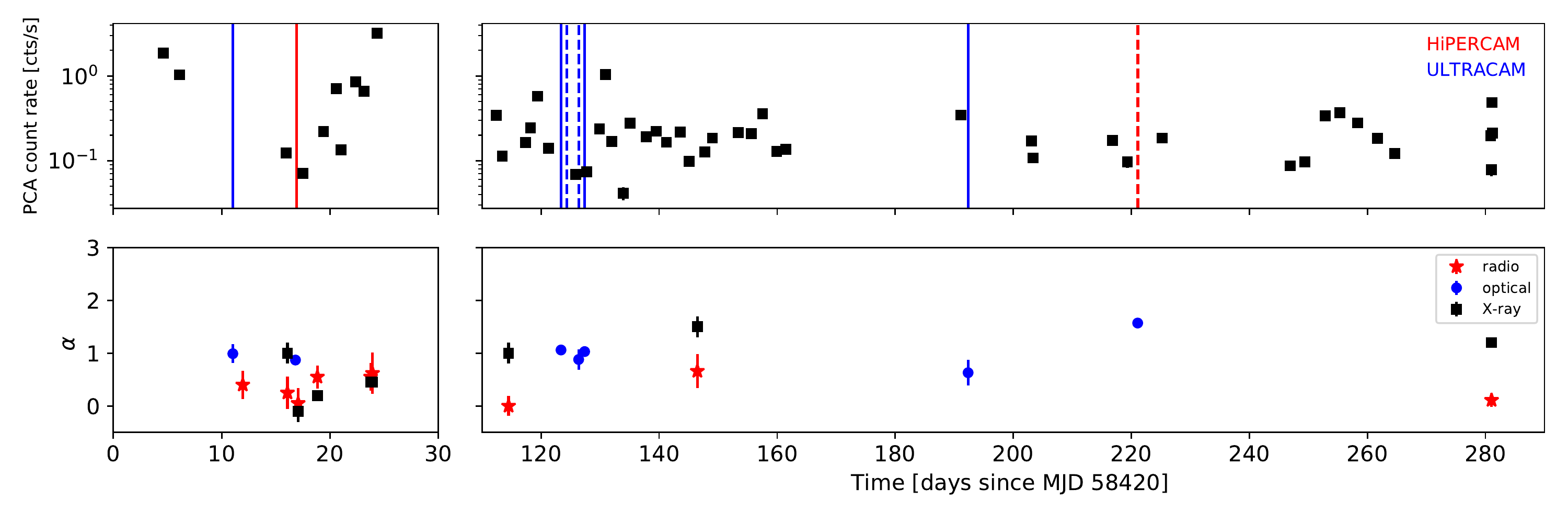}
\caption{The long-term X-ray light curves and the radio and X-ray spectral index light curves of \target\ during its 2018 and 2019 outburst. The top panel shows the Swift/XRT PC mode X-ray data (black squares), where the vertical lines mark the times of our ULTRACAM (blue) and HiPERCAM (red) optical observations. The dashed lines show the times when the optical observations were simultaneous with NICER. The bottom panel show the radio (red stars; 4.5\,GHz) and X-ray (black squares; 0.2--10\,keV) spectral indices ($F_\nu \propto \nu^\alpha$) taken from \citet{Eijnden20}. The blue squares show the spectral index of the optical flares observed with ULTRACAM and HiPERCAM determined in this paper (see Section\,\ref{sec:properties}). }
\label{fig:longterm}
\end{figure*}

\section{Introduction}
\label{sec:intro}

The low-mass X-ray binary \target\ was discovered as an X-ray transient on 2018 October \citep{Krimm18} with the Burst Alert Telescope (BAT) aboard the Neil Gehrels {\it Swift} Observatory \citep{Gehrels04}. Subsequent multi-wavelengths observations  detected the source at longer wavelengths. The Ultraviolet and Optical Telescope ({\it Swift-UVOT}) on-board  {\it Swift}  detected a variable UV source which was  coincident with a previously detected UKIRT Infrared Deep Sky Survey (UKIDSS) and Pan-STARRs source \citep{Kennea18}. Optical follow-up observations revealed that the source had brightened by $\sim$2.5 magnitudes \citep{Vasilopoulos18}. The source was also detected in the radio by the Arcminute Microkelvin Imager Large Array having a variable flux density of 300--600\,$\mu$Jy at 15.5\,GHz \citep{Bright18}. At X-ray wavelengths, the outburst was relatively faint, with a flux of $\sim$10$^{-11}$\ergs\  at 0.5--10\,keV and a hard spectrum with a photon index of $\Gamma$ = 2 \citep{Reynolds18}.

Superimposed on the outburst were bright, short X-ray flare events \citep{Ludlam18,Hare19} where the observed flux increased by more than an order of magnitude in a few seconds \citep{Hare20}. Optical flares were also identified \citep{Vasilopoulos18, Baglio18, Rajwade18, Rajwade19, Paice18} with wavelength-dependent optical variability on time-scales of minutes, and sporadic, fast `red' flares on  time-scales of seconds \citep{Paice18}. The timing characteristics were reminiscent of those seen in the black hole X-ray binary V404\,Cyg, which showed long-term `blue' flaring and short-term sporadic `red' flaring during its 2015 outburst \citep{Kimura16, Gandhi16}. The radio emission from \target\ showed variability by up to a factor of $\sim$8 on  time-scales of minutes due to mass accretion rate fluctuations consistent with a compact jet \citep{Bright18, Eijnden20}. The X-ray spectrum  showed evidence for significant intrinsic local absorption \citep{Reynolds18, Hare20} and the P-Cygni profile observed in the optical spectrum \citep{Munoz20}, suggested that a significant amount of mass was ejected from the inner accretion flow. 

Although \target\ entered the Sun constraint for most X-ray telescopes in 2019 November,  it was detected again with the Monitor of the All-sky X-ray Imager (MAXI) in 2020 February in a previously unobserved X-ray state, with  significantly less variability and enhanced soft X-ray emission, implying  a transition to a soft state  \citep{Negoro20, Buisson20a}. During 2020 March several Type I X-ray bursts were detected with the Neutron star Interior Composition Explorer (NICER) and the Nuclear Spectroscopic Telescope Array (NuSTAR), identifying \target\ as a neutron star binary system despite the fact that pulsations were not  detected \citep{Buisson20_burst}. These bursts exhibited photospheric radius expansion allowing a distance estimate of $\sim$12.8 kpc. Strong periodic drops in X-ray flux were also detected, consistent with eclipses by the secondary star and variable obscuration due to the thickness of the disc/accretion stream  which is also responsible for the  strong variability \citep{Buisson21}.

Here we report on high time-resolution HiPERCAM and ULTRACAM optical observations of \target\ some of which are simultaneous with NICER observations, taken in 2018 and 2019. We comment on the observed optical flaring and on the optical/X-ray flux correlations and timing properties of the light curves.

\section{Observations}

In Fig.\,\ref{fig:longterm} we show the long-term X-ray light curve of \target\ during its 2018 and 2019 outburst and mark the optical and X-ray observations presented in this paper.

\subsection{NICER -- X-rays}
\label{sec:obs:X-rays}

\target\ was observed with NICER in an intensive monitoring program during its 2018 and 2019 X-ray outburst. NICER is an X-ray instrument on board the International Space Station (ISS) where individual photons with energies in the range 0.2--12\,keV can be detected with a time resolution of 40\,ns \citep{Gendreau16}. The data reduction was carried out using the collection of NICER-specific tools \textsc{nicerdas} which is part of HEASARC \footnote{\url{https://heasarc.gsfc.nasa.gov}}. Full Level2 calibration and screening was conducted with \textsc{nicerl2}, which calibrated, checked for good time intervals, merged, and cleaned the data. The barycentric correction was carried out using \textsc{barycorr}, and finally the photon events were binned to the times of the optical light curves as described in the following sections. We produced a light curve in the 0.2--12\,keV energy band for each data segment using \textsc{xselect} and then applied the background correction.  In order to calculate the hardness ratio, we extracted light curves in the 0.5--3.0\,keV and 3--10\,keV bands. For these  light curves, we normalised each incoming photon with respect to the effective area of the telescope at that energy. We define the hardness ratio of the X-rays as (hard-soft)/(hard+soft), where the hard and soft X-ray rates are in the 3--10\,keV and 0.5--3.0\,keV range, respectively. The errors on the hardness ratio were calculated by using 1-$\sigma$ Poisson errors (following the example of \citealt{Gehrels86}) to simulate maximum and minimum values of the individual X-ray bands, and then calculating the hardness ratio at each extreme. We note that these errors are an approximation only and may underestimate any outliers.

%
\begin{figure*}
\subfloat[The ULTRACAM light curves]
{\includegraphics[width=0.74\textwidth]{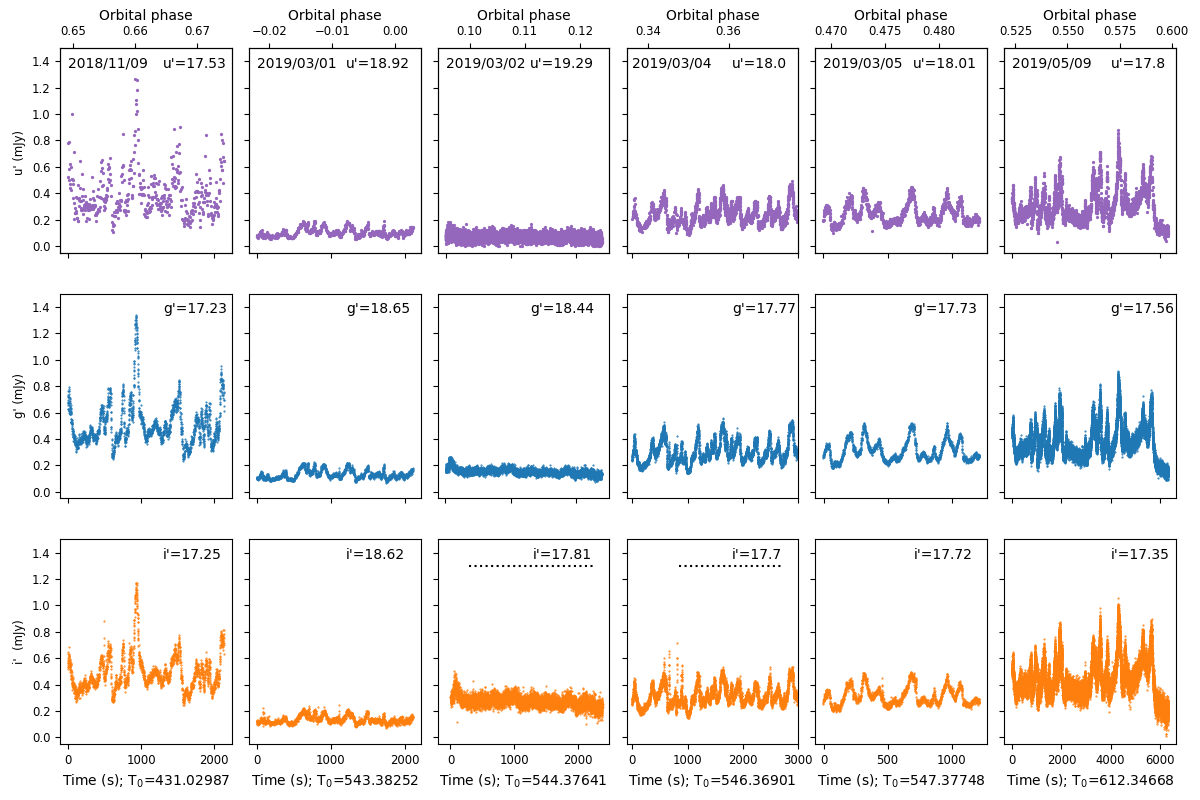}
\label{fig:lcurves_ucam}}

\subfloat[The HiPERCAM light curves]
{\includegraphics[width=0.71\textwidth]{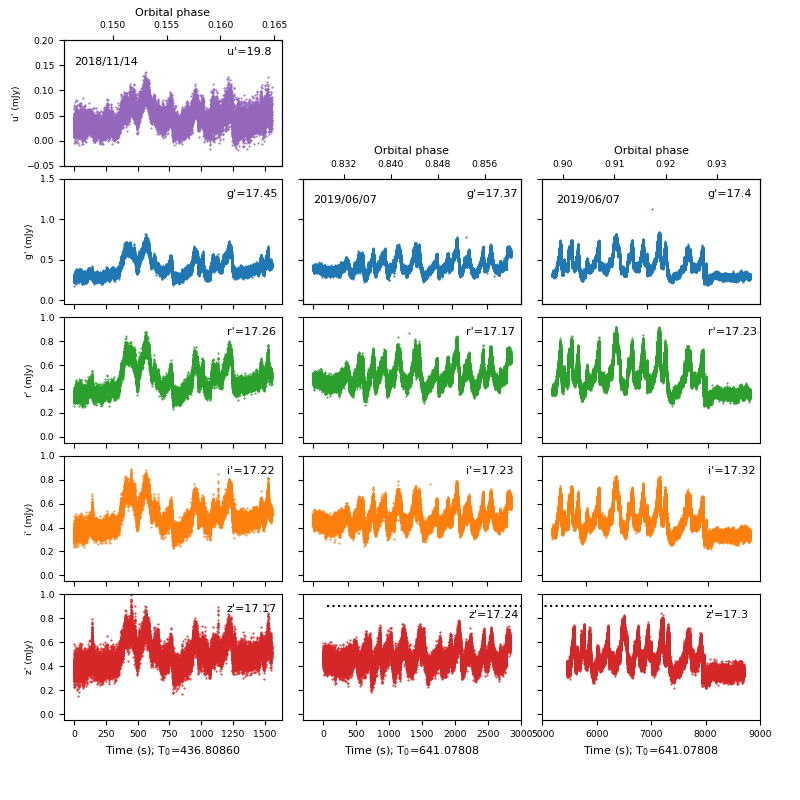}
\label{fig:lcurves_hcam}}
\caption{The observed ULTRACAM  (top) and  HiPERCAM (bottom) light curves of \target. The black dotted horizontal line shows the time of NICER observations. The mean magnitude of \target\ is shown in each panel. A MJD time offset of $T_0$ + 58000.0 ($T_0$ is in days) is applied and we use the orbital ephemeris given in \citet{Buisson21}.}  
\end{figure*}

\subsection{ULTRACAM/NTT -- Optical}
\label{sec:obs:ucam}

High-speed  multi-colour photometry of \target\ was carried out using  ULTRACAM instrument \citep{Dhillon07} on the 3.5\,m New Technology Telescope (NTT) in La Silla, Chile. ULTRACAM uses dichroic beamsplitters to simultaneously image three custom made Sloan Digital Sky Survey (SDSS) filters, and can observe at frame-rates well above 100\,Hz due to the frame-transfer CCDs and the lack of a physical shutter \citep{Dhillon07}. We used ULTRACAM to observe \target\ during 2018 November, 2019 March and 2019 May. The 2018 observations were carried out simultaneously with the $u'$, $g'$, and $i'$  SDSS filters \citep{Doi10}, whereas the 2019 observations were performed using the higher throughput $u_s$, $g_s$, and $i_s$  Super-SDSS filters \citep{Dhillon21} which use multi-layer coatings rather that coloured glass to define the filter bandpasses, with the cut-on/off wavelengths designed to match higher throughput the original SDSS filters. Unlike most observations of this type, the times were not explicitly chosen to coincide with X-ray observations. Some of the observations did overlap with the X-ray observations performed with the NICER instrument and such simultaneity was purely serendipitous (see Section\,\ref{sec:obs:X-rays}). On different nights, ULTRACAM was used in windowed mode (one window containing the target and the other containing multiple comparison stars)  with 1$\times$1 binning. Typically, compact binaries are faint in the $u_s$-band, and so ULTRACAM's on-chip co-adding feature was used, which provides a longer exposure time in the $u_'$-band so as  to increase the signal-to-noise ratio. The details of the observing setup for each night are given in Table\,\ref{table:log}. 

We used the HiPERCAM pipeline software\footnote{\url{https://github.com/HiPERCAM/hipercam}} to debias, flat-field and extract the target count rates using aperture photometry with a seeing-dependent circular aperture tracking the centroid of the source. The sky background was computed using the clipped mean of an annular region around the target and relative photometry of \target\ was carried out with respect to the local standard star (PSO\,J185832.982-081400.913). For the $r_s$-band and $g_s$-band, the field is covered by the Pan-STARRS survey and so the calibrated $r_s$-band and $g_s$-band magnitudes are listed in DR1 catalog \citep{Magnier20}. These were transformed  to SDSS magnitudes \citep{Finkbeiner16} and then used to calibrate the target light curves. Since the field is not covered by any archival optical survey in the $u_s$-band, calibrating these data was less straightforward. Flux standards were observed on various nights during the ULTRACAM observations in 2019 March. These flux standards were used to determine the  $u_s$-band instrument zero-point. The local standards were then calibrated which in turn were used to calibrate the target light curve. For the nights when no flux standard was observed, we assume that the  $u_s$-band  zero-point measured during the March observing runs was still valid. The difference between the ULTRACAM Super-SDSS and SDSS filters leads to an uncertainty in the flux calibration of $<$ 3 per cent \citep{Wild22}. The observed ULTRACAM light curves are shown in Fig.\,\ref{fig:lcurves_ucam}. 

\subsection{HiPERCAM/GTC -- Optical}
\label{sec:obs:hcam}

Sub-second optical imaging was carried out in 2018 November and  2019 June  using HiPERCAM on the 10.4\,m Gran Telescopio Canarias (GTC) in La Palma, Spain. HiPERCAM uses dichroic beamsplitters to simultaneously image the custom made Super-SDSS $u_s$, $g_s$, $r_s$, $i_s$ and $z_s$ filters. Similar to ULTRACAM, HiPERCAM can  observe at frame-rates well above 1000\,Hz which is achieved by the lack of a physical shutter and the frame-transfer CCDs that can rapidly shift charge into a storage area for reading out, freeing up the original pixels for observation and thereby achieving low (7.8\,ms) dead-times\,\citep{Dhillon21}. The CCDs were binned by a factor of 4 and drift mode was used with four windows (336$\times$200 pixels each) for all the observations. The instrument was orientated so that one window was centered on \target\ and another window on a local standard star. We used an exposure time of 43.6\,ms and 44.9 which resulted in a cadence of 46.6\,ms and 47.9\,ms, for the 2018 and 2019 observations, respectively (see Table\,\ref{table:log} for details). Observations were obtained on two nights, 2018 November 14 and 2019 June 7. The observations taken in 2019 were coordinated with the X-ray instrument NICER. A log of the observations is given in Table\,\ref{table:log}. Similar to the ULTRACAM data, we used the HiPERCAM pipeline software to debias, flat-field and extract the photon counts for the target and local standard using aperture photometry with a seeing dependent circular aperture. The local standard stars used are  listed in the Pan-STARRS survey DR1 catalog \citep{Magnier20} and have $g'$, $r'$, $i'$ and $z'$ magnitudes which were transformed  to SDSS magnitudes \citep{Finkbeiner16} and then used for the photometric calibration of \target. For the 2018 data  the $u'$-band calibration was determined using the local standard star PSO\,J185827.968-081329.815 and the full-frame acquisition images which was calibrated by determining the instrument zero-point.  As a check we also determined the $g_s$, $r_s$, $i_s$ and $z_s$ magnitudes and found that they agreed with the Pan-STARRS magnitudes at the $<$10 per cent level. Unfortunately, the local standard star PSO\,J185826.795-081357.216 used in the 2019 observations was not detected in the $u_s$-band images, and so it could not be flux calibrated. The difference between the HiPERCAM Super-SDSS and SDSS filters leads to an uncertainty in the flux calibration of $<$ 3 per cent \citep{Brown22}. Finally we convert from SDSS magnitudes to flux density, where we propagate the uncertainty in the local standard. The observed HiPERCAM light curves are shown in Fig.\,\ref{fig:lcurves_hcam}.

\section{Reddening}
\label{sec:ebv}

\target's position in the sky allows us to estimate the line of sight interstellar reddening. The Galactic neutral atomic hydrogen (H \textsc{I}) column density to the target is $N_{\rm H} \sim 1.84 \times 10^{21}$\,cm$^{-2}$ \citep{HI4PI16}. Using the relation between the Galactic hydrogen absorption column density and optical extinction \citep{Foight16} along with the galactic extinction law \citep{Cardelli89} we determine a colour excess of $E\left(B-V\right)$=0.21 mag. We can also estimate $N_{\rm H}$ from spectral fits to the NICER data.  Using the XSPEC \citep{Arnaud96} software package, a blackbody and power-law model fit to the 2018 November data gives $N_{\rm H} \sim 2.0-2.5 \times 10^{21}$\,cm$^{-2}$, whereas fits to the 2019 June data gives $N_{\rm H} \sim 1.6-1.7 \times 10^{21}$\,cm$^{-2}$. The value for $N_{\rm H}$ determined from the NICER data is consistent with the value determined from the H \textsc{I} maps and we assume a colour excess of 0.21 mag for the rest of this paper.

\begin{figure}
\centering
\includegraphics[width=0.9\linewidth,angle=0]{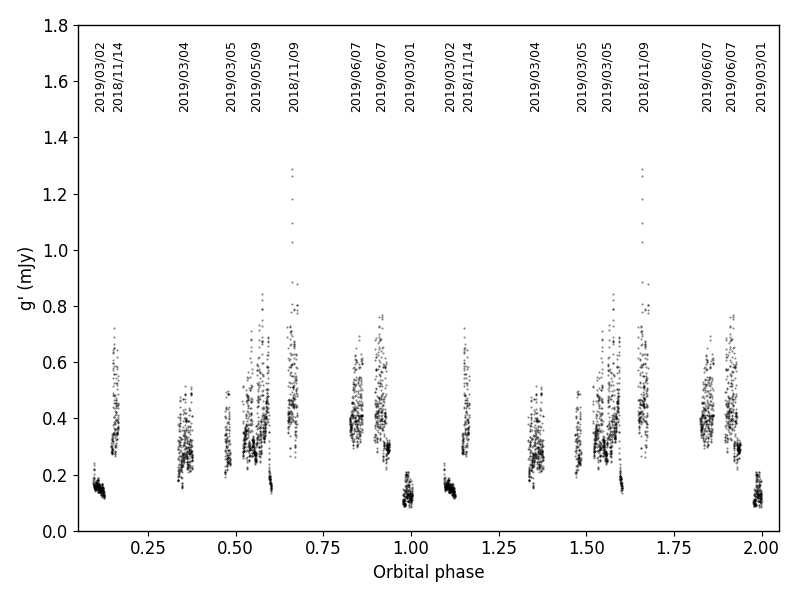}
\caption{ULTRACAM and HIPERCAM $g_s$-band light curves of \target\ as a function of orbital phase using the  orbital ephemeris given in \citet{Buisson21}, where phase 0.0 is defined as superior conjunction of the compact object. For clarity two orbital phases are plotted and the light curves have been rebinned to a time resolution of 10\,s.}
\label{fig:phase_lc}
\end{figure}

\section{Optical flares}
\label{sec:properties}

In Figs.\,\ref{fig:lcurves_ucam} and \ref{fig:lcurves_hcam} we show the observed HiPERCAM and ULTRACAM light curves, respectively, where wavelength dependent flaring activity is clearly seen. Flaring is superimposed on a sinusoidal modulation, which is due to a combination of the secondary star's ellipsoidal modulation, X-ray heating and other possible sources of light in the system (see Fig.\,\ref{fig:phase_lc}). To determine the properties of the flares first use the colour excess of $E\left(B-V\right)$=0.21 mag determined in Section\,\ref{sec:ebv} with the interstellar extinction law \citep{Cardelli89} to deredden the observed fluxes. We identify and isolate the flare events by determining the start and end of the same flare event in each waveband.  
We then subtract the interpolated flux underneath the flare event which in effect subtracts the contribution of the non-variable component.
We assume during the actual flare event that the other components that contribute to the observed flux do not vary. We define small and large flares as events with  $g_s$-band amplitudes of $\sim$0.1\,mag and $\sim$1\,mag, respectively. A total of 102 large and 5 small flare events were isolated, respectively.  Fig.\,\ref{fig:flares_examples} of the Appendix shows some examples of the isolated flare events where flares on different time-scales, amplitude and colour are clearly seen. For the flare events we also determine the peak flare flux in each waveband and flux ratio.

In Fig.\,\ref{fig:phase_lc} we show the observed $g_s$-band light curve of \target\ as a function of orbital phase, using the orbital ephemeris given in \citet{Buisson21}  where phase 0.0 is defined as superior conjunction of the compact object.  Although our orbital phase coverage is relatively poor ($\sim$33 per cent), we observe flares at all orbital phases. \citet{Buisson21} find that the bright flares occur preferentially in the post-eclipse phase of the orbit, around orbital phase $\sim$0.3, most likely due to increased thickness at the disc-accretion stream. We do not find any evidence for this in our optical data, but note our poor phase coverage. We find that the mean flux and the intrinsic source fractional RMS variability defined as  $\sigma^2_{\rm source}$ = $\sigma^2_{\rm total}$ - $\sigma^2_{\rm noise}$ \citep{Vaughan03} are strongly linearly correlated with a Pearson's correlation coefficient of 0.84. The low RMS observed at phase 0.0 (2019 March 01) which has the lowest flux of our observations and very little flaring is consistent with a system at a high binary inclination angle \citep{Buisson21,Knight22}.

\begin{figure}
\includegraphics[width=1.0\linewidth,angle=0]{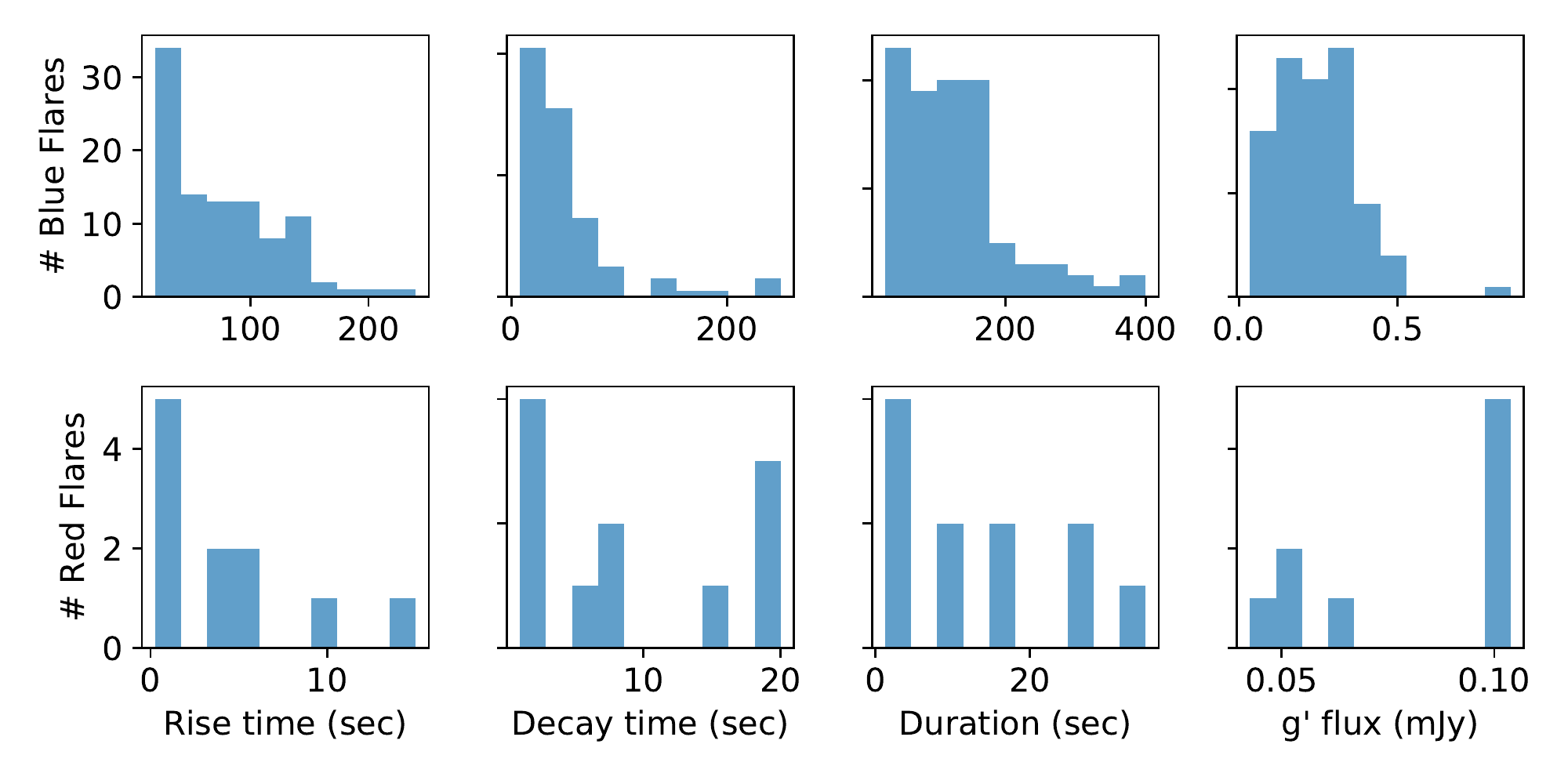}
\caption{The rise, decay time-scales, duration and flux histograms of the dereddened flare events. } 
\label{fig:hist}
\end{figure}

\begin{figure}
\includegraphics[width=1.10\linewidth,angle=0]{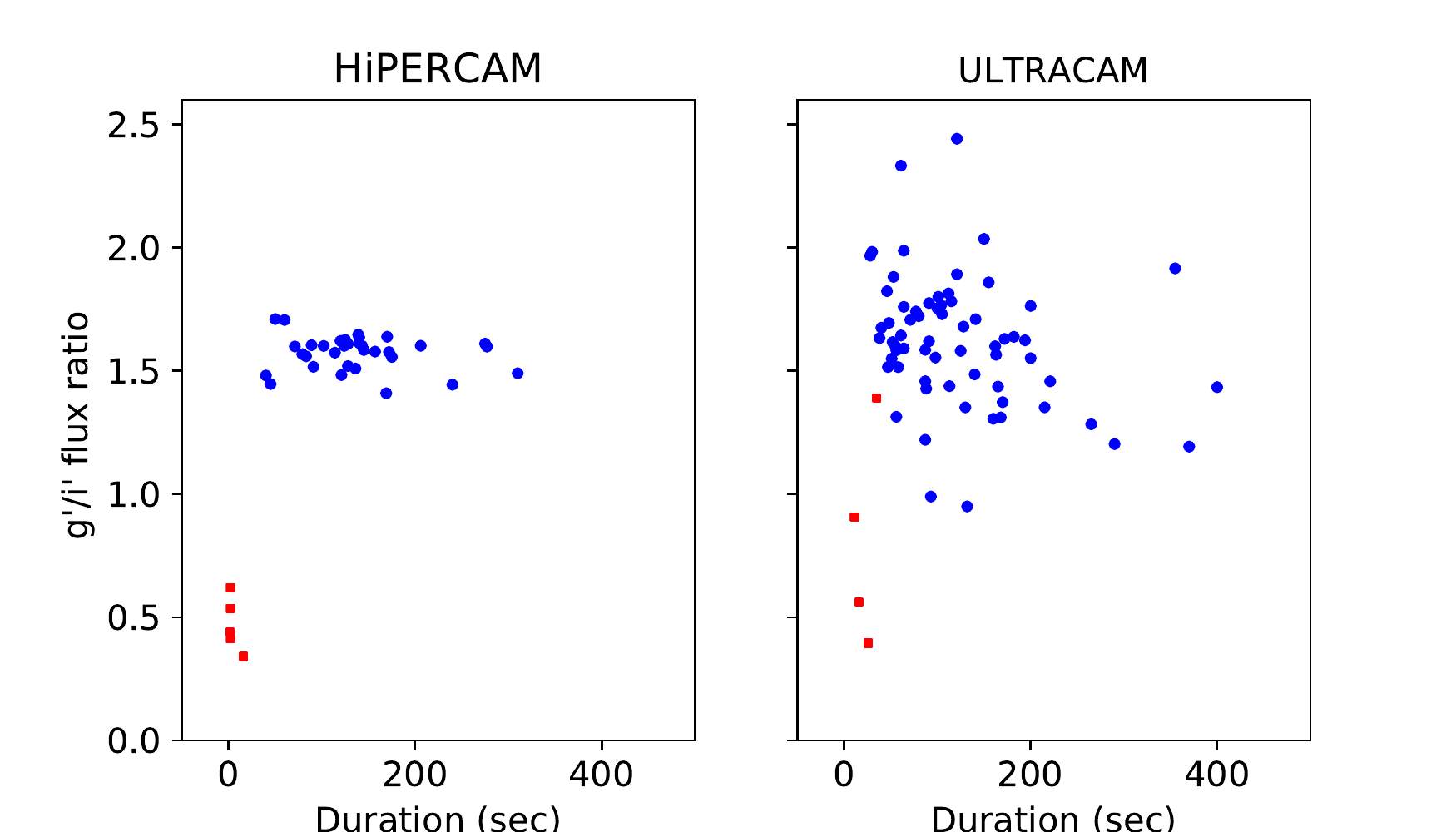}
\caption{The flare duration versus colour of the dereddened small (red points) and large flare (blue points) events.}
\label{fig:amp}
\end{figure}

\begin{figure}
\centering
\includegraphics[width=0.9\linewidth,angle=0]{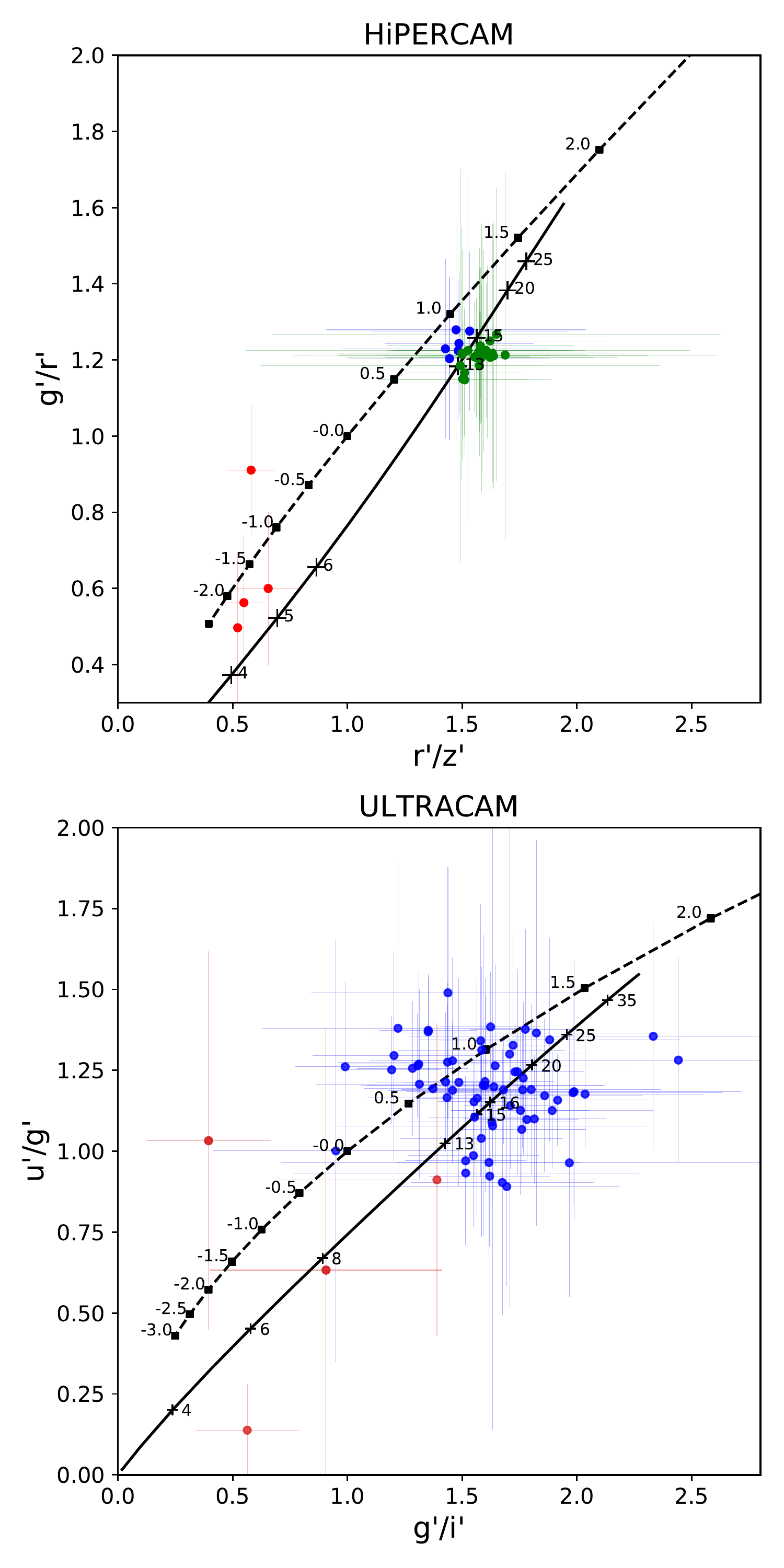}
\caption{Colour--colour diagram for the ULTRACAM and HiPERCAM large and small dereddened flare events. The dashed line shows a power-law model of the form $F_\nu \propto \nu^\alpha$, where the black squares mark the value of  $\alpha$ ranging from  -2.0 to  +2.0 in units of 0.5. The solid black line is a blackbody model where the crosses show the  temperature in units of 1000\,K.  In the top panel the red  circles show the HiPERCAM small flares, whereas the blue  (2018 November 14) and green circles (2019 June 7) show the HiPERCAM large flare events. In the bottom panel the red and blue circles show the ULTRACAM small and large flares events, respectively.}
\label{fig:colcol}
\end{figure}

\subsection{Time-scales}
\label{sec:time-scale}

We determine  the rise, decay and duration of the dereddened flares which are shown in Fig.\,\ref{fig:hist}. As one can see, the `red' flares (more flux at longer wavelengths) have a much shorter time-scale and amplitude compared to the `blue' flares  (more flux at shorter wavelengths) events.  The `red` and `blue' flares have median  $g_s$-band amplitudes of $\sim$0.1\,mag and $\sim$1\,mag, respectively. In Fig.\,\ref{fig:amp} we show the flare duration versus colour. The flares are separated into two regions: short-duration `red' flares and long-duration `blue' flares. Small amplitude 'red` flares are observed on 2018 November 14 (HiPERCAM) and 2019 February 2 (ULTRACAM), whereas large `blue' flares are present in all observations, except on 2019 March 2 where no flares are observed. The different time-scales and amplitudes of the flares indicate that they arise from different emission processes.

\subsection{Spectral energy distribution}
\label{sec:sed}
 
In an attempt to interpret the broad-band spectral properties of the flares, we compare the observed fluxes with the prediction for different emission mechanisms, namely  synchrotron emission and blackbody. The latter has an approximately power-law form on the Rayleigh--Jeans tail and so we characterise the synchrotron and blackbody emission with a power-law form $F_\nu \propto \nu^\alpha$, where $\nu$ is the frequency and $\alpha$ is the spectral index. We compute the given emission spectrum and then calculate the expected flux density ratios in the relevant filters using the synthetic photometry package \textsc{synphot} in \textsc{iraf/stsdas}. For the blackbody emission, given the intrinsic model flux we then determine the corresponding radius of the region that produces the observed dereddened flux at a given distance.

In Fig.\,\ref{fig:colcol} we show the HiPERCAM and ULTRACAM  individual peak flare flux ratios and the expected results for different emission models.  We show the $g_s$, $r_s$ and $z_s$ fluxes common to the HiPERCAM 2018, 2019 and ULTRACAM 2019 data sets and the  $u'$, $g'$ and $i'$ fluxes for the ULTRACAM 2018 data set. Fig.\,\ref{fig:flares_sed} of the Appendix shows some example  fits to the individual dereddened flare events observed on 2018 November 14 (HiPERCAM) and 2019 May 9 (ULTRACAM). The power-law indices obtained by fitting the broad-band spectral energy distribution of the individual large and small flare events are in the range $\alpha \sim $-1.0 to -2.0 (with a mean of $\alpha \sim$ -1.5) for the `red' flares. In contrast the `blue' flares can be represented with a power-law  of $\alpha \sim$ 1.0 (range of $\alpha \sim $0.6 to 1.2) or a $\sim$14,000\,$\pm$\,2000\,K blackbody which with a mean $g_s$ peak flare flux of $\sim$0.45\,mJy (out of eclipse) corresponds to a radius of  $\sim$1.0$\,\pm$\,0.2\,\Rsun, assuming a distance of 12.8\,kpc \citep{Buisson20_burst}. Although a single temperature blackbody  has limited physical significance and is likely a very poor description of a flare event, it is useful for comparison with other works. The 2019 data was taken at orbital phase $\sim$ 0.9 which is outside the start of eclipse ingress \citep{Buisson21} and so we can rule out a decrease in $N_{\rm H}$ due to the  absorption in the atmosphere of the secondary star. However,  \citet{Castro22} have detected disc winds in the hard state and the associated variable obscuring columns that contribute to $N_{\rm H}$ might explain the differences we observe.


\begin{figure*}

\subfloat[ULTRACAM/NICER 2019 March 2]
{\hspace*{-5mm}\includegraphics[width=0.45\textwidth]{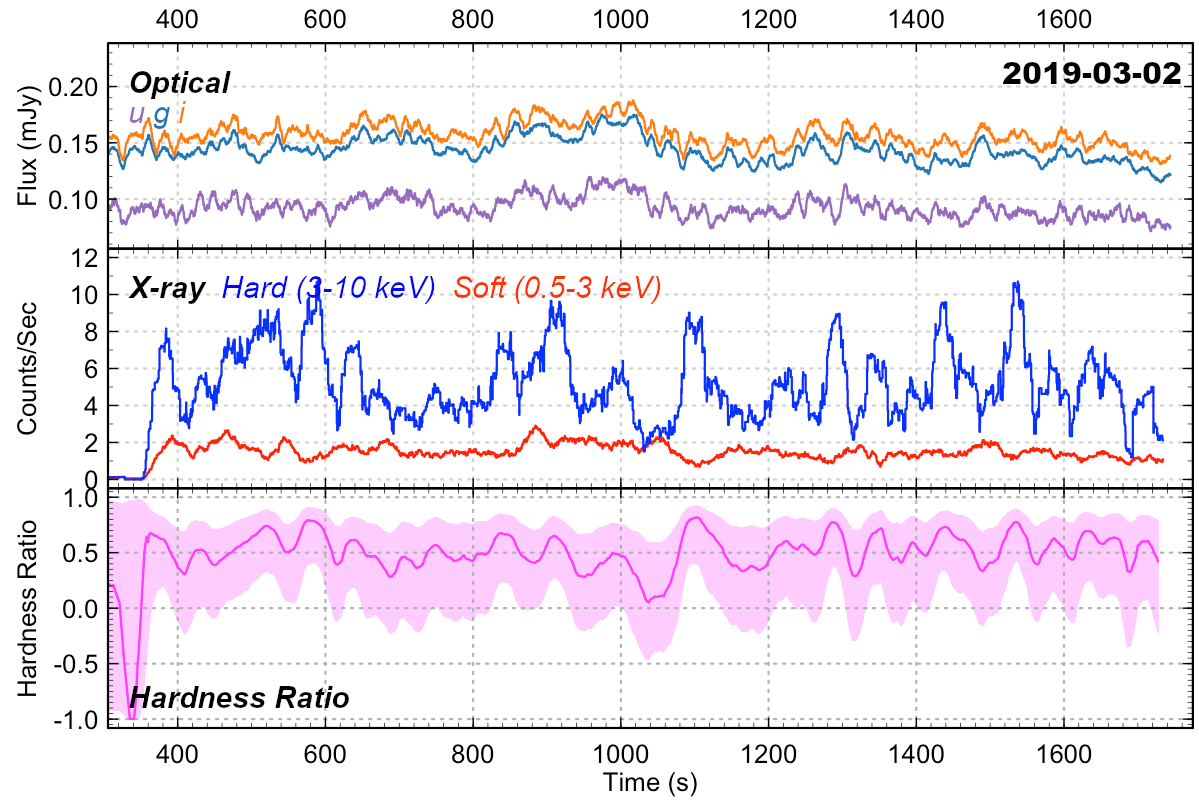}}
\subfloat[ULTRACAM/NICER 2019 March 4]
{\hspace*{+5mm}\includegraphics[width=0.45\textwidth]{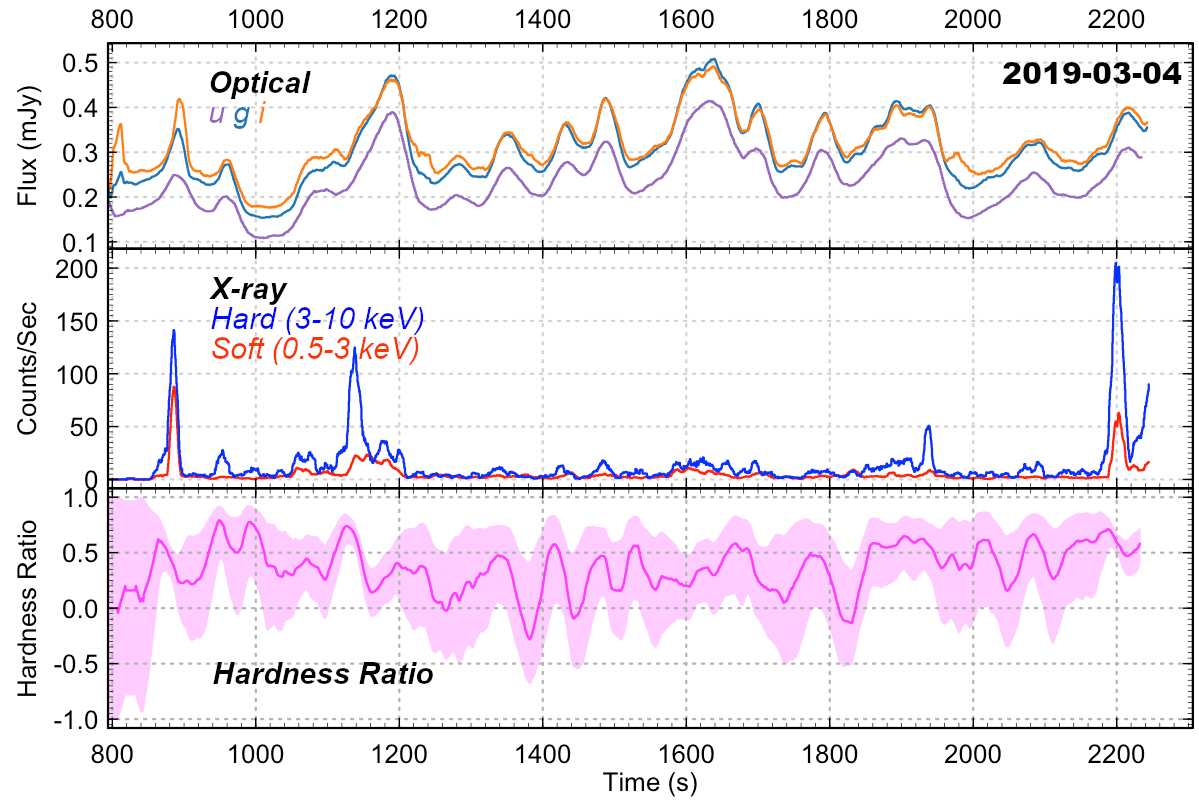}}

\subfloat[HiPERCAM/NICER 2019 June 7; part1]
{\hspace*{-5mm}\includegraphics[width=0.45\textwidth]{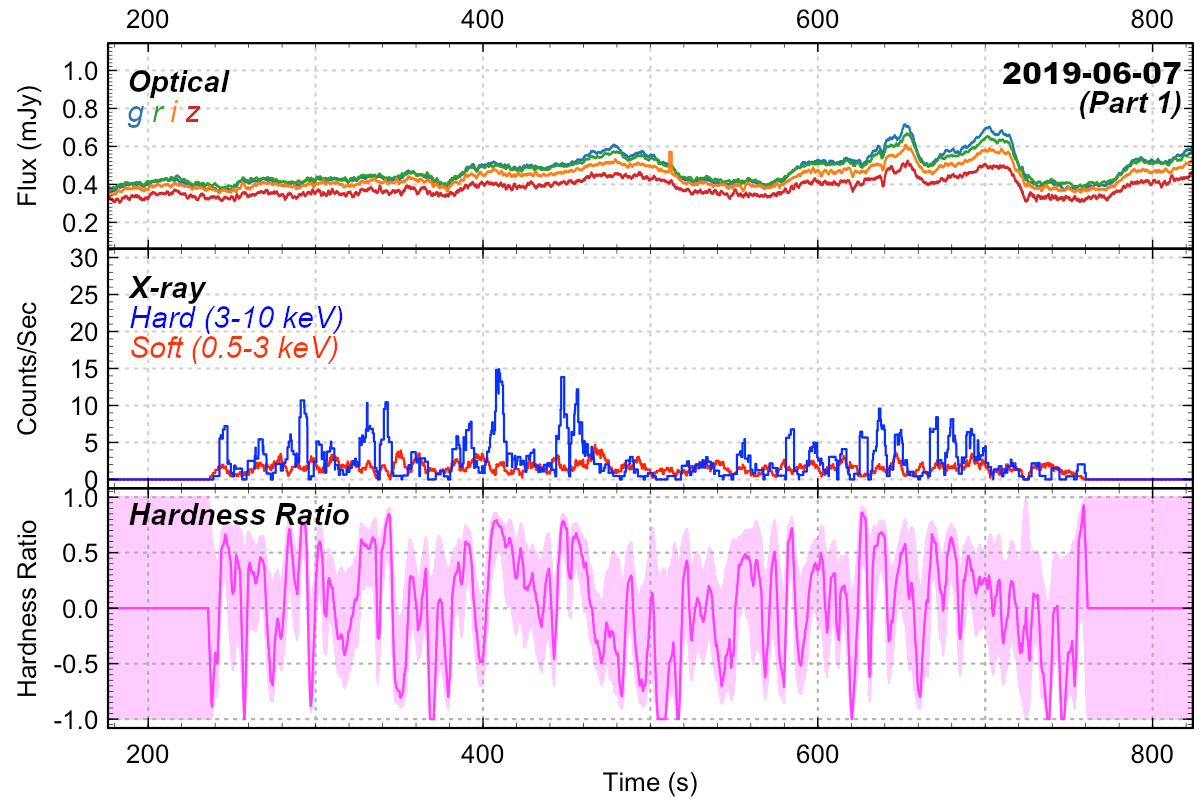}}
\subfloat[HiPERCAM/NICER 2019 June 7; part2]
{\hspace*{+5mm}\includegraphics[width=0.45\textwidth]{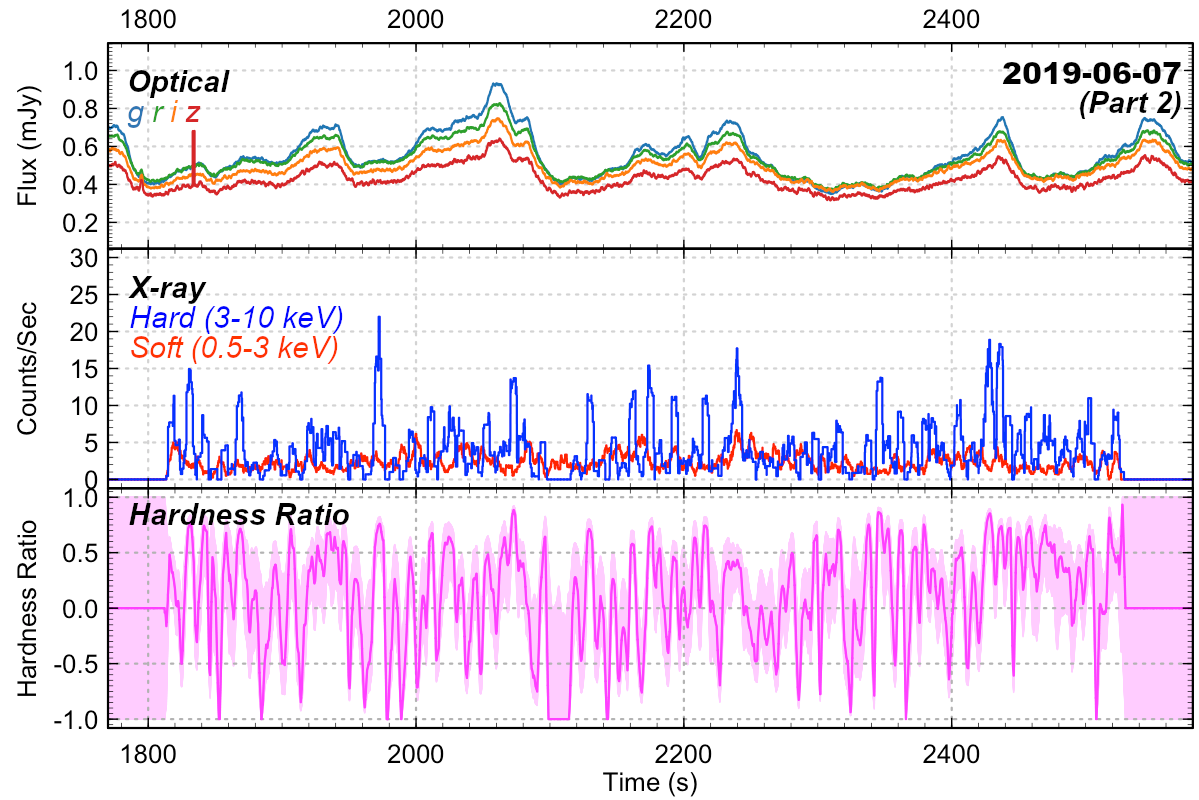}}

\subfloat[HiPERCAM/NICER 2019 June 7; part3]
{\hspace*{-5mm}\includegraphics[width=0.45\textwidth]{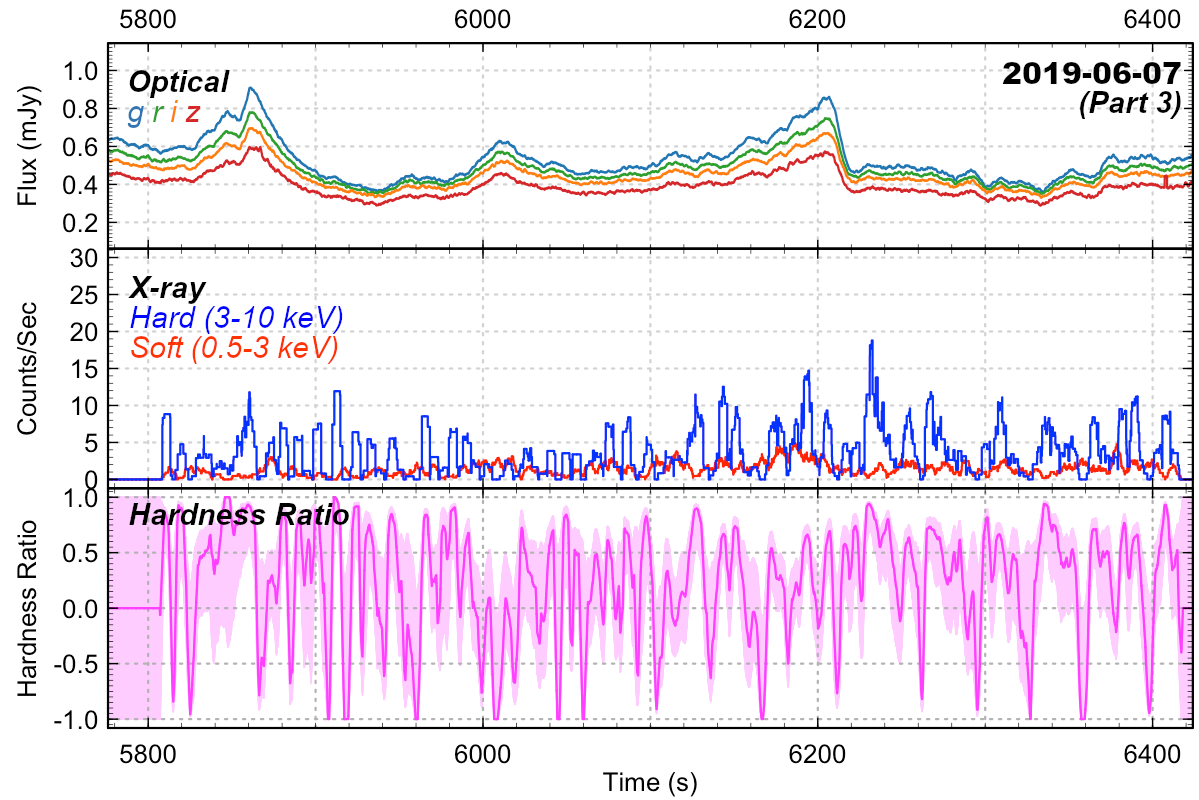}}
\subfloat[HiPERCAM/NICER 2019 June 7; part4]
{\hspace*{+5mm}\includegraphics[width=0.45\textwidth]{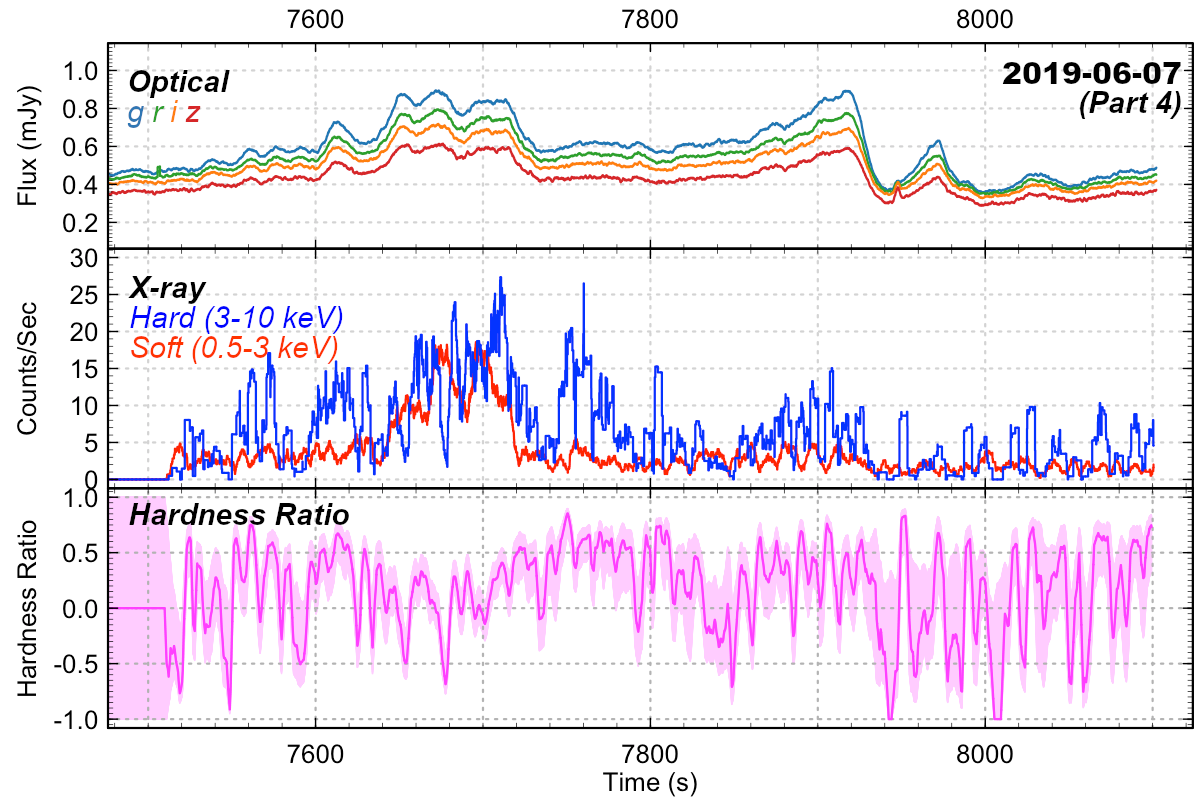}}

\caption{The simultaneous optical and X-ray light curves of \target. From top to bottom, the optical light curves, the hard (3--10\,keV; blue) and soft (0.5--3.0\,keV; red) X-ray light curves and the X-ray hardness ratio defined as the ratio of the rates (hard-soft)/(hard+soft). The X-rays and hardness ratio light curves have been binned with a moving average of 100 points for readability (except for 2019 March 4 where a 20 point moving average was used due to the much higher count rates). A barycentered MJD time offset of 58544.37641329, 58546.36898883 and 58641.08379497 has been applied to the 2019 March 2, March 4 and June 7 data, respectively.}
\label{fig:lc} 
\end{figure*}


\begin{figure*}

\subfloat[ULTRACAM/NICER 2019 March 2]
{\includegraphics[width=0.5\textwidth]{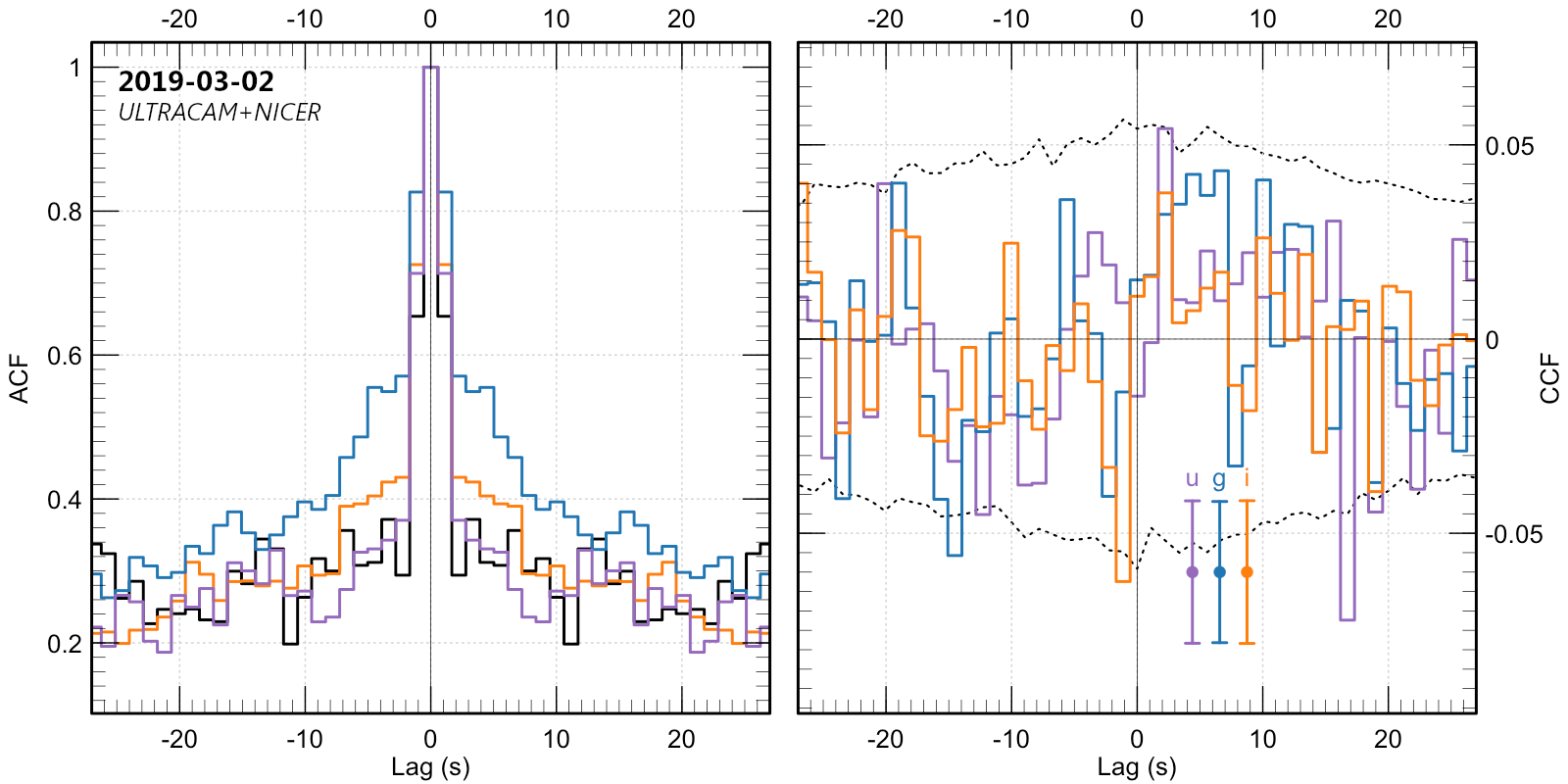}}
\subfloat[ULTRACAM/NICER 2019 March 4]
{\includegraphics[width=0.5\textwidth]{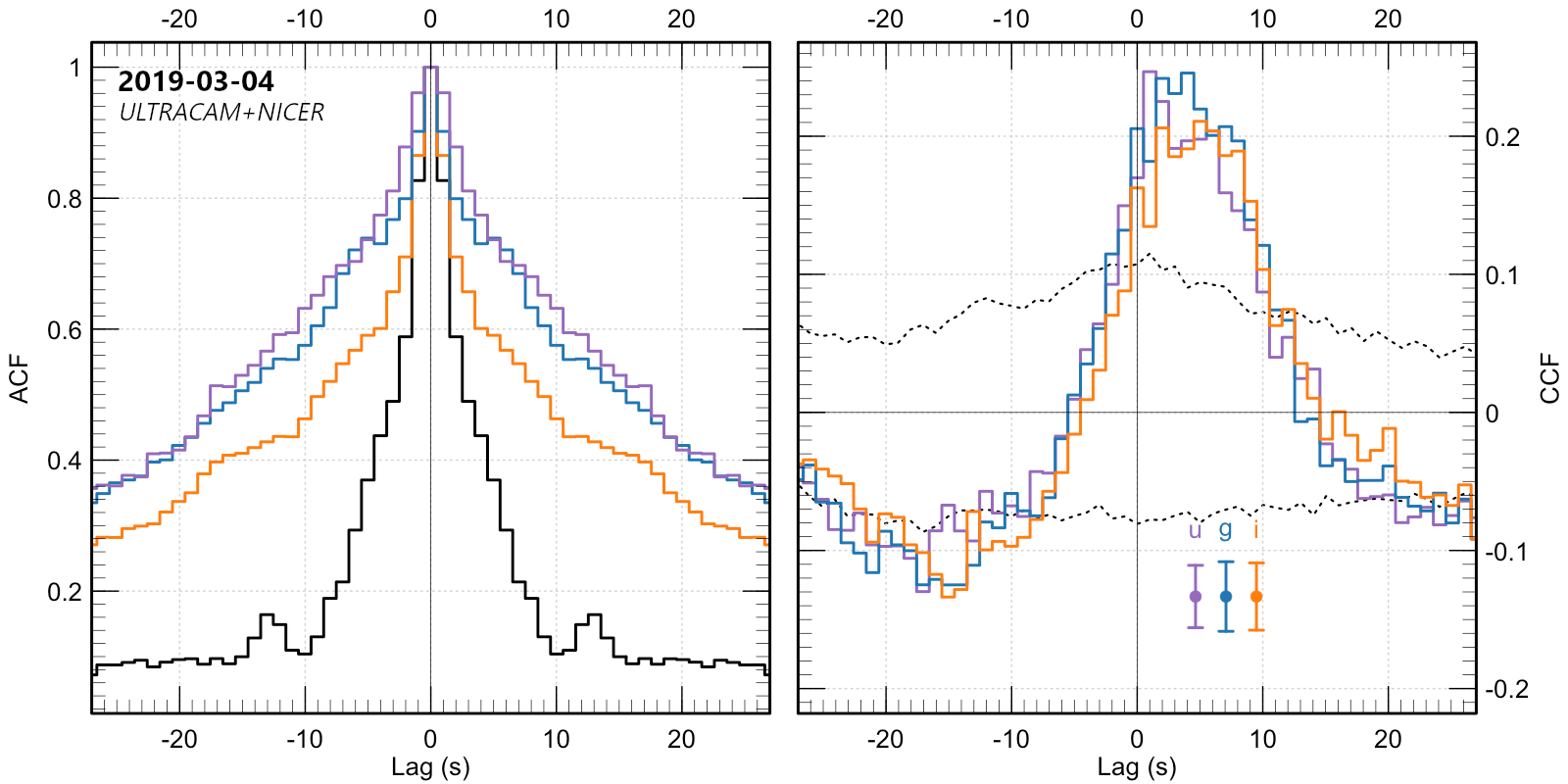}}

\subfloat[HiPERCAM/NICER 2019 June 7; part1]
{\includegraphics[width=0.5\textwidth]{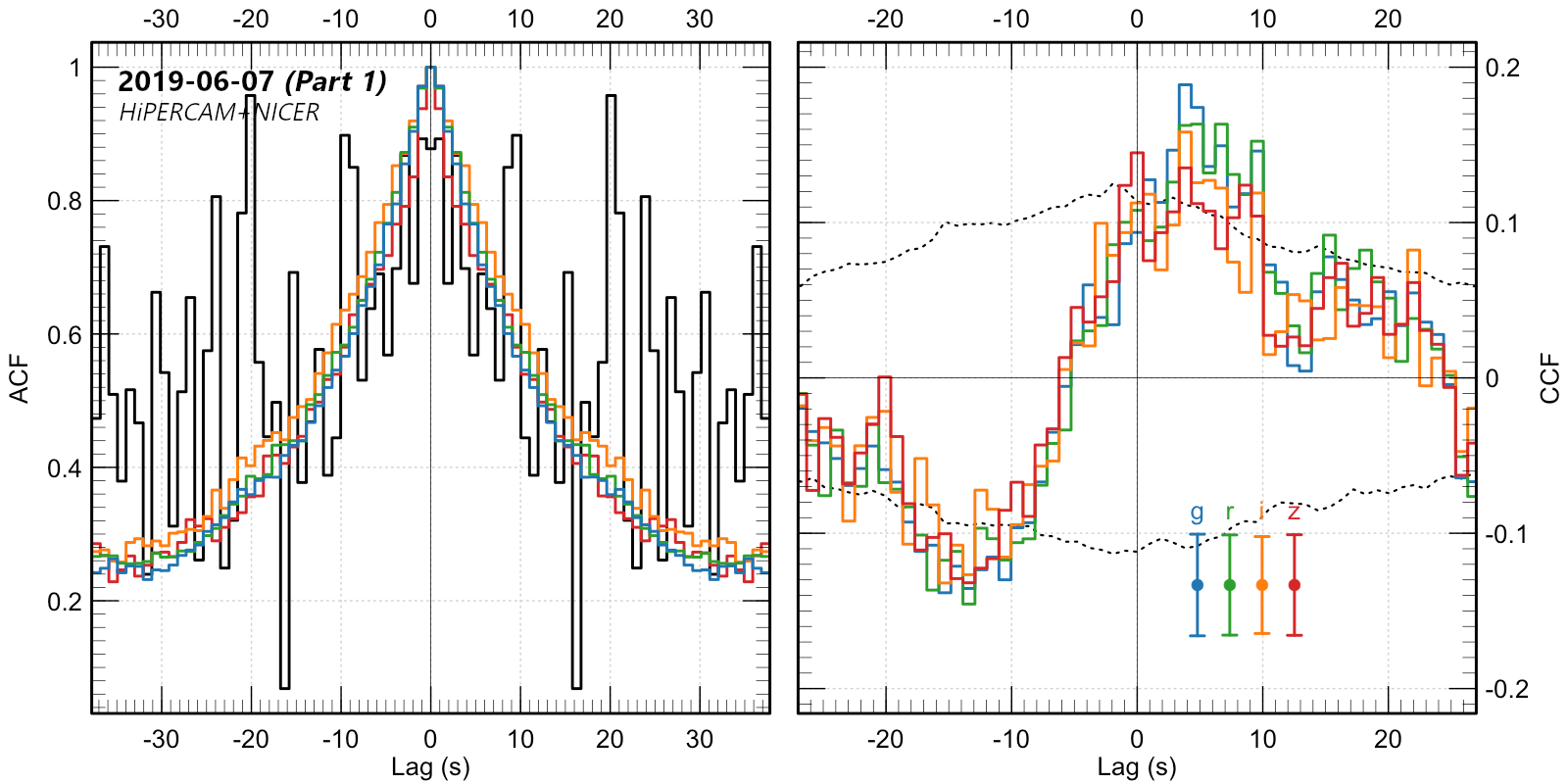}}
\subfloat[HiPERCAM/NICER 2019 June 7; part2]
{\includegraphics[width=0.5\textwidth]{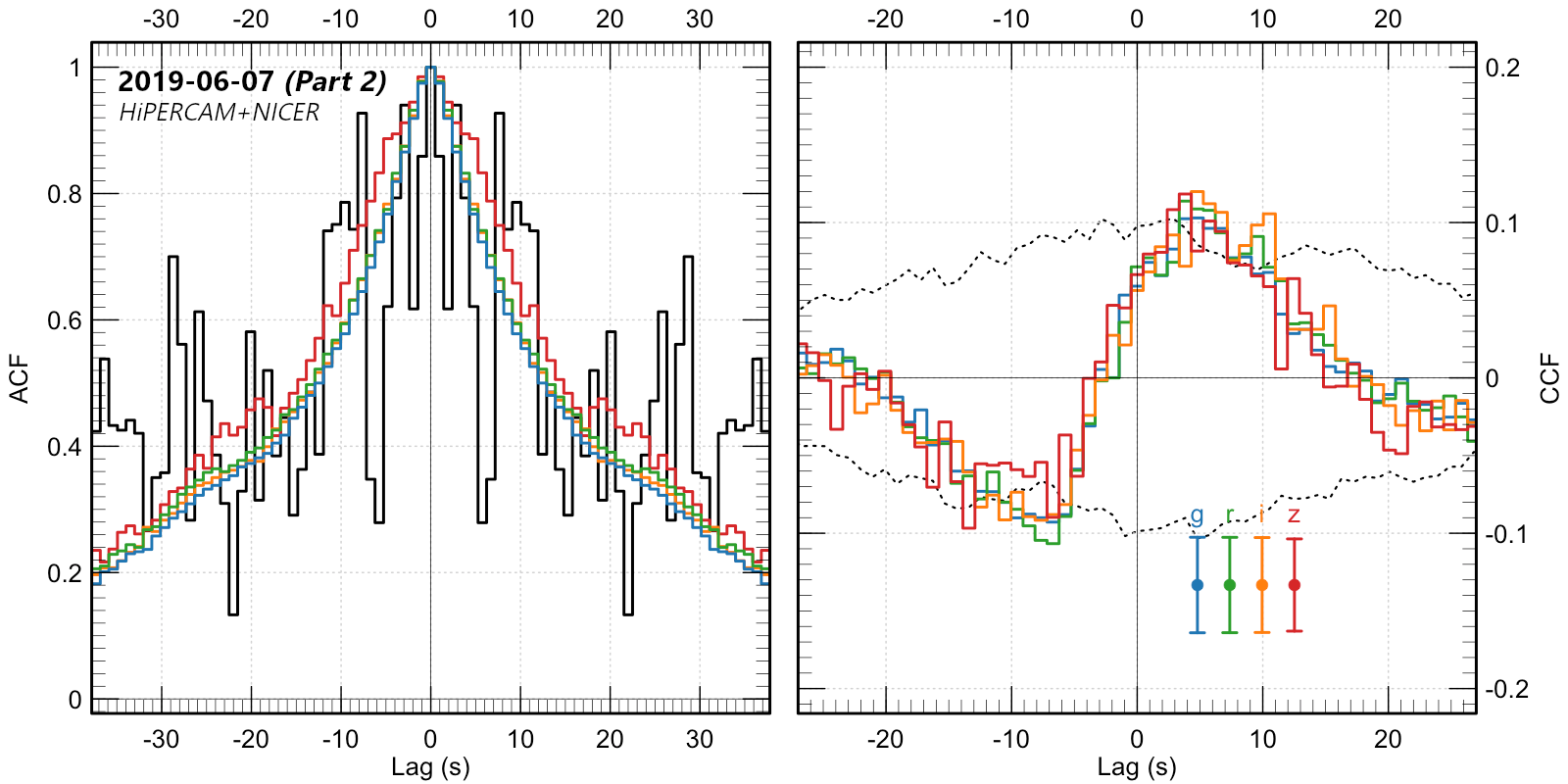}}

\subfloat[HiPERCAM/NICER 2019 June 7; part3]
{\includegraphics[width=0.5\textwidth]{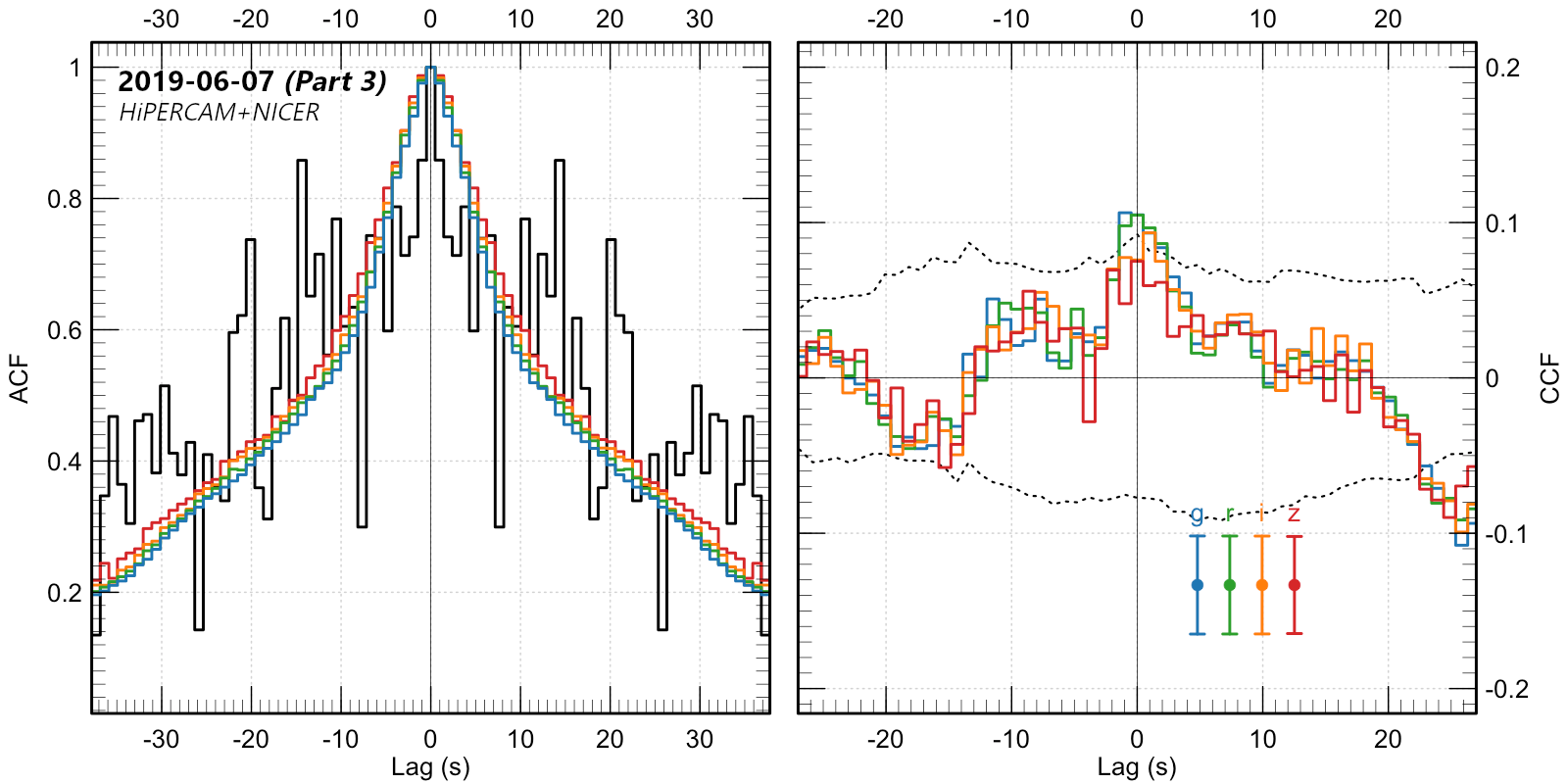}}
\subfloat[HiPERCAM/NICER 2019 June 7; part4]
{\includegraphics[width=0.5\textwidth]{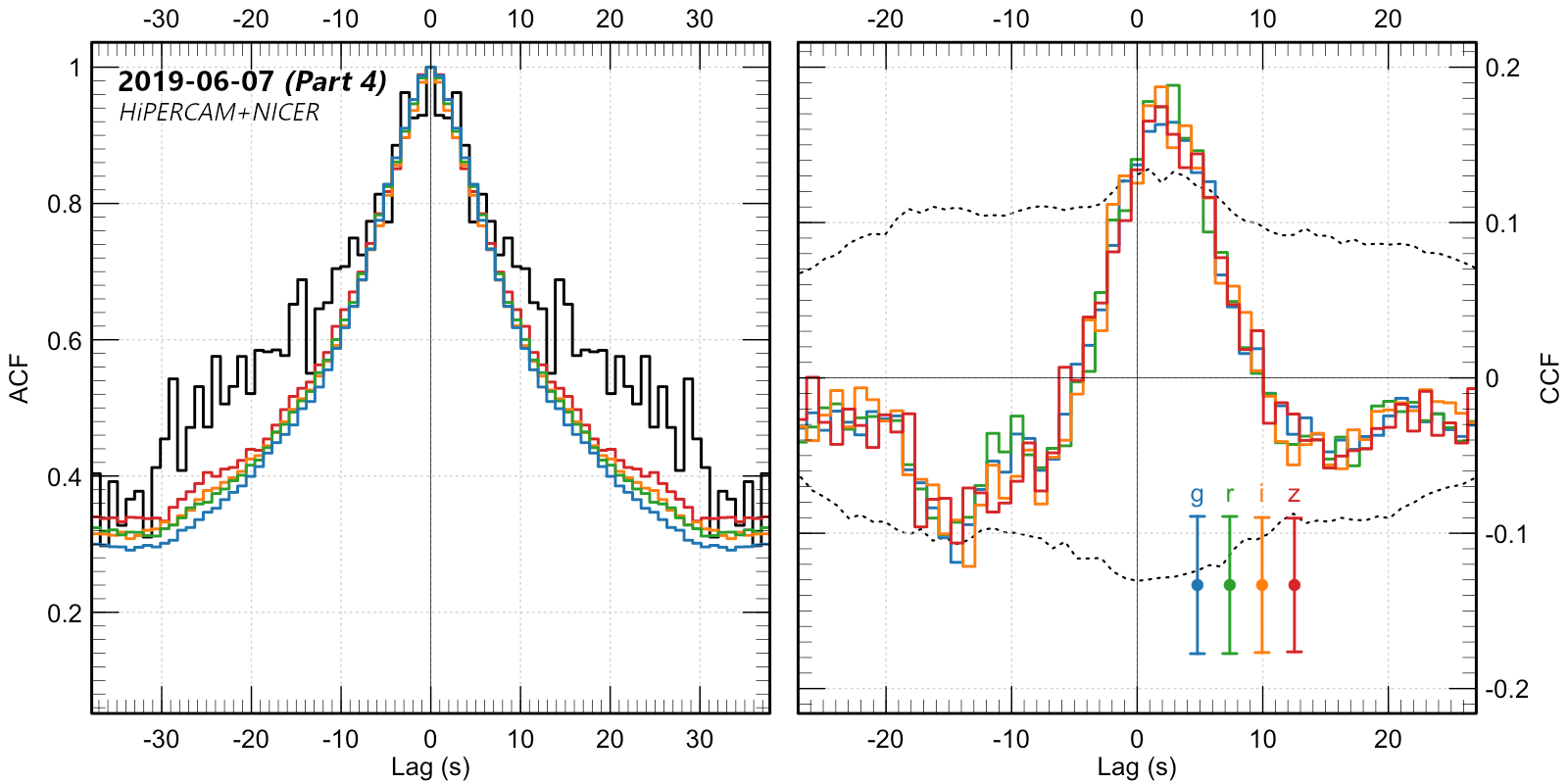}}

\caption{The ACF (left plot) and CCF (right plot) of the simultaneous optical and X-ray light curves of the 2019 March 2 (a), 2019 March 4 (b) and 2019 June 7 (c to f) data. A positive lag implies that the optical flux lags the X-ray flux. For the 2019 June 7 data (c to f) we show  the corresponding ACFs and CCFs of the data split into four sections, corresponding to the sections when the data were simultaneous. In the left panel, the ACF of the X-ray data is shown in black  and the ACF of the $g_s, r_s, i_s,$ and $z_s$ data are shown in blue, green, orange and red, respectively. In the right panel the CCF of the X-ray data with respect to the $g_s, r_s, i_s$ and $z_s$ data are in shown blue, green, orange and red, respectively. The black dashed lines represent the 5 and 95 percent confidence intervals.}
\label{fig:acf_ccf} 
\end{figure*}

\section{Timing and correlation analysis}
\label{sec:timing}

The auto-correlation function (ACF) analysis of the individual optical and X-ray light curves and the cross-correlation function (CCF) of the simultaneous optical and X-ray light curves can also be used to constrain the emission processes and location, respectively. We perform such a timing analysis on the simultaneous optical and X-ray data using the same methods/techniques  outlined in \citet{Paice19}. We use the NICER X-ray light curves and the dereddened ULTRACAM and  HiPERCAM optical light curves determined in Sections\,\ref{sec:obs:X-rays} and \ref{sec:properties}, respectively. To create the simultaneous light curves we first corrected the times of both datasets to the solar system barycentre and then binned the X-ray photons directly to the optical time bins. Since the optical light curves have a constant dead-time, the X-ray photons observed during these times are not used. For the 2019 June 7 HiPERCAM data, we show the four different simultaneous sections, whereas for the ULTRACAM dataset we show the two simultaneous sections taken in 2019 March 2 and 4.

\subsection{Optical/X-ray correlations}

In Fig.\,\ref{fig:lc} we show the simultaneous optical and X-ray light curves taken on 2019 March 2, 4 and June 7. For the X-ray data we also show the  hardness ratio of the X-ray count rates. The CCF shows the response of the optical light curves to variations in the X-ray light curve as a function of time lag. Positive time lags  indicate a net correlation in which the optical flux lags the X-ray flux. The CCF is produced by splitting and detrending the simultaneous light curves into segments of equal length. We determine the CCF for each segment and calculate the mean CCF and standard error in each bin. We also compute the auto-correlation functions (ACFs) of the X-ray/optical light curves. The Poisson noise dominating the X-ray ACFs at zero lag is  corrected by making use of the Wiener--Khinchin theorem, which states that the power spectrum of a random process and its ACF are Fourier pairs. Therefore, we can subtract the white noise from the X-ray power spectrum and then compute the inverse Fourier transform to determine the ACF. In Fig.\,\ref{fig:acf_ccf} we plot the corresponding ACF and CCFs for all our simultaneous optical/X-ray light curves.  
To determine the confidence levels in the CCFs we simulate 1000 similar (yet uncorrelated) optical light curves, compute the cross-correlation function with respect to the X-ray light curve and then determine the the 5 and 95 percent boundaries in each bin of the CCF lag. We create the optical light curves by first computing the Fourier transform of the optical light curve, randomising the arguments and then performing the inverse Fourier transform to create a lightcurve with an identical power spectrum. In the following,for each simultaneous dataset we summarise the observed characteristic of the light curves and average ACFs and CCFs. 

\begin{itemize}

\item
For the 2019 March 2 data the mean X-ray count rate is 2.6 counts\,s$^{-1}$ over the length of the simultaneous ULTRACAM observation. Low optical and X-ray variability is observed with no significant flaring behaviour, compared to what is observed on other nights. In general the X-ray light curve has a strong hard component. The optical ACF is broader than the  X-ray ACF which is what one expects if the optical flux arises from X-ray reprocessing. No significant features are observed in the CCFs.

\item
For the 2019 March 4 data, the mean count rate is relatively high at 7.9 counts\,s$^{-1}$ over the length of the simultaneous ULTRACAM observation. A few relatively strong X-ray flare events are observed which have a strong hard component. The optical ACF is broader than the  X-ray ACF, consistent with X-ray reprocessing. One can clearly see that the  optical and X-ray fluxes are correlated, which provides a visual confirmation of the CCF observed.  The CCF of this observation shows the strongest positive correlation of any of our epochs, with a peak at a time lag of $\sim$5\,s in every band (a coefficient of $\sim$0.3 is a significant value in fast-timing studies of X-ray binaries; see e.g. \citealt[][]{Gandhi10, Gandhi17, Paice19}). A weak negative correlation at negative lags is also seen at $\sim$-5\,s. Furthermore, there appears to be a repeated phenomenon in the light curves - the hard X-rays increase first and then give way to softer X-rays.  This is more clearly seen in the flare at time $\sim$1100 \,s. In the CCFs, there appears to be a correlation between the optical delay and wavelength, in which the $u_s$-band delay is shorter than the $g_s$-band delay, which is shorter than the $i_s$-band delay. This implies that reprocessing is dominant.

\item
Finally, the 2019 June 7 data has a relatively low mean count rate of 0.65 counts\,s$^{-1}$ coincident with the HiPERCAM observations. Although the X-ray variability is much lower, several optical peaks do have slight increases in X-ray count rates, where the increase seems to be slightly greater in the hard X-rays. In general the X-ray light curve has a hard component but slightly softer than other epochs and is dominated by a large flare event in part 4 at  7700\,s, which has a strong soft component as noted by the change in the X-ray hardness ratio. This is in contrast to the other short-term X-ray flare events which seem to have a hard component. The ACF and CCF properties in sections 1 to 3 are very similar. The parts 1 to 4  data show the optical ACF and X-ray ACF to be similar in shape. A relatively strong positive correlation in the CCF with a peak at a time lag of  $\sim$5\,s is observed in every band for parts 1, 2 and 4, and at a time lag of $\sim$0\,s in the part 3 data. A weak negative correlation at negative lags is also observed between $\sim -$20\,s and $-$10\,s.

\end{itemize}

\subsection{Optical/X-ray correlations of flaring Events}

\begin{figure}
\includegraphics[width=0.49\textwidth]{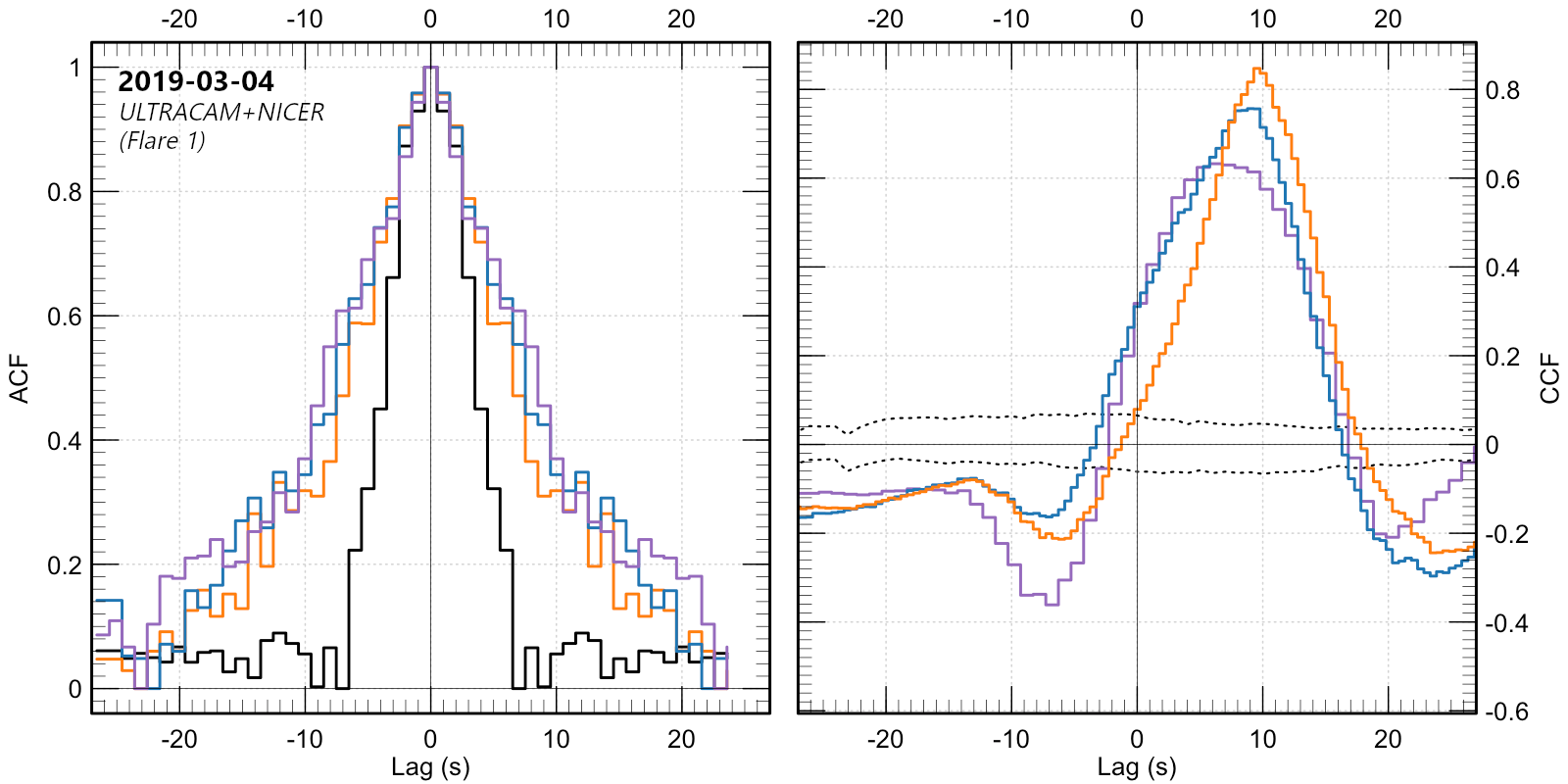}

\includegraphics[width=0.49\textwidth]{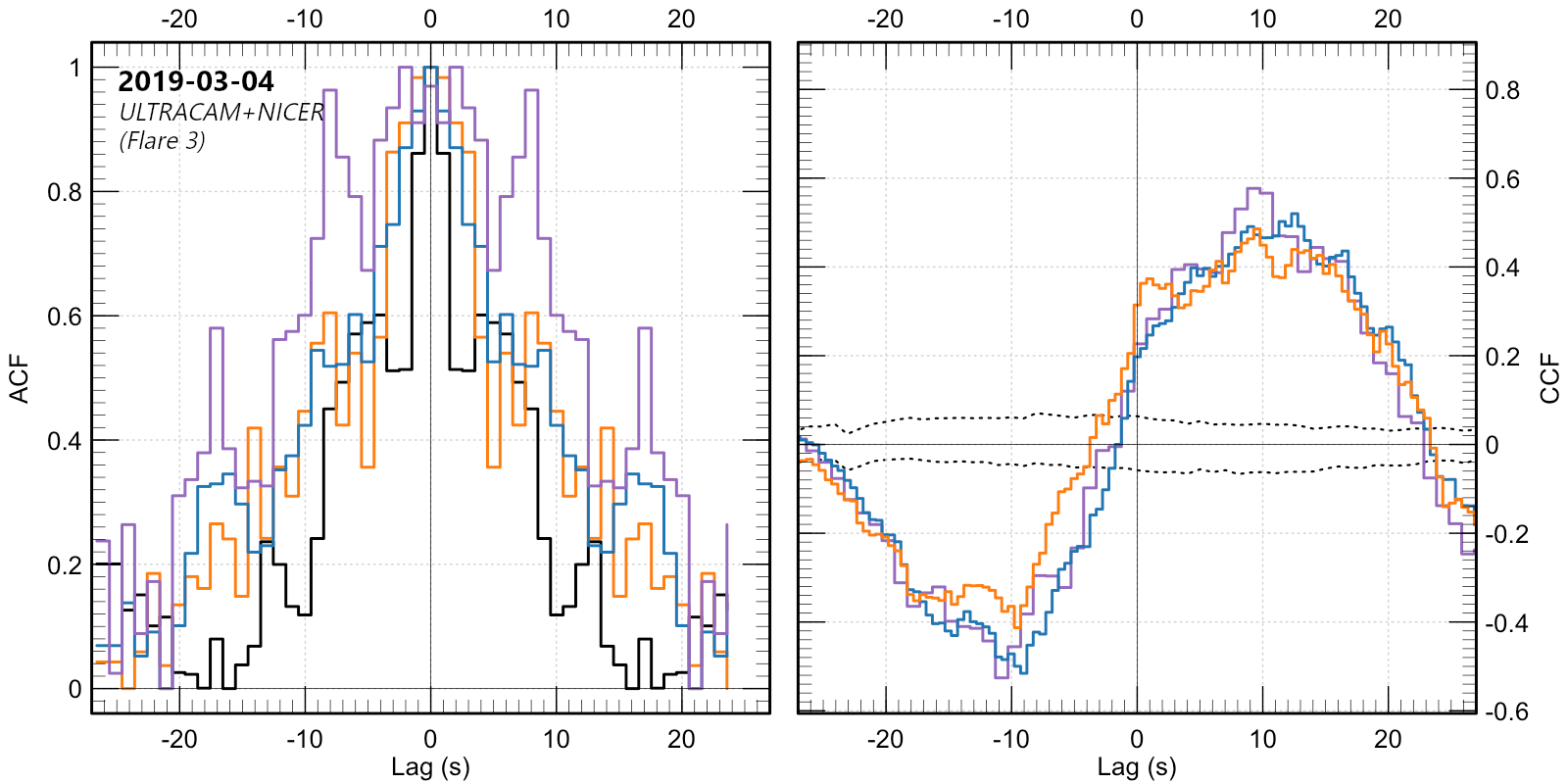}

\includegraphics[width=0.49\textwidth]{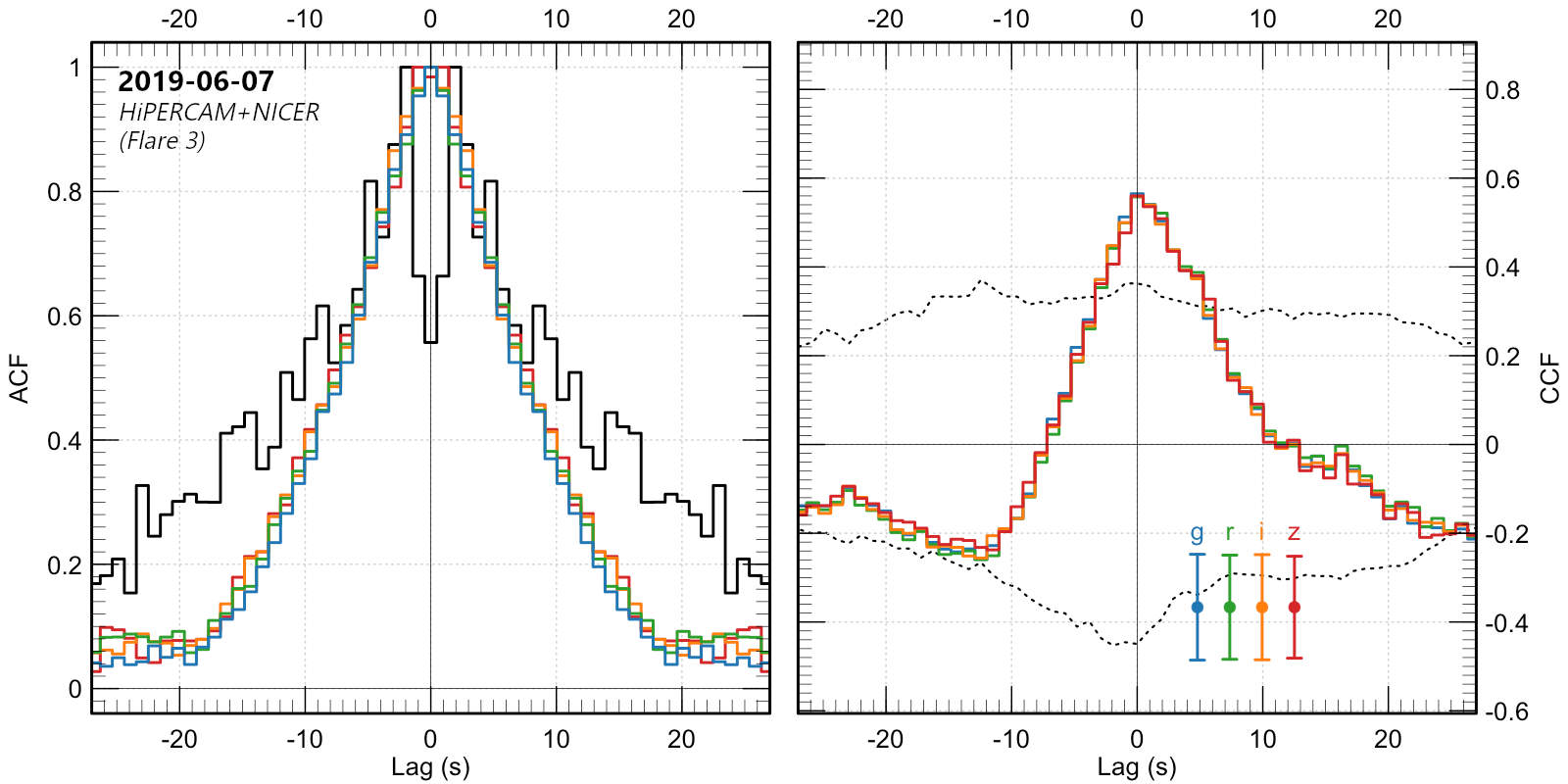} 
\caption{Same as Fig.\,\ref{fig:acf_ccf} but for the clear flare events on 2019 March 4 
and 2019 June 7.}
\label{fig:acf_ccf_flare} 
\end{figure}

In order to further investigate the flaring events we determine the ACFs and CCF for three clearly defined flare events on 2019 March 4. We compute the optical and X-ray ACFs and well as the optical/X-ray CCF using a 100\,s window (see Fig.\,\ref{fig:acf_ccf_flare}). As one can see, the CCFs of the flare events share many characteristics, including a high CCF correlation (0.4--0.8) with lags between 0--5\,s and a precognition dip. The 2019 March 4 and 2019 June 9 data are taken at orbital phase $\sim$0.35 and $\sim$0.93, respectively. Indeed, the flare events taken at different orbital phases have time delays consistent with arising from reprocessing in the secondary star. One expects the longest time delay to arise at orbital phase quadrature (phase 0.25) and the shortest at superior conjunction of the secondary star (phase 0.0). Indeed, if one had sufficient flare events across the binary orbit one could perform echo-mapping in order to extract the fundamental binary parameters \citep{O'Brien02}.

\begin{figure*}
\hspace{0mm}\subfloat[ULTRACAM/NICER 2019 March 2]
{\includegraphics[width=0.31\textwidth]{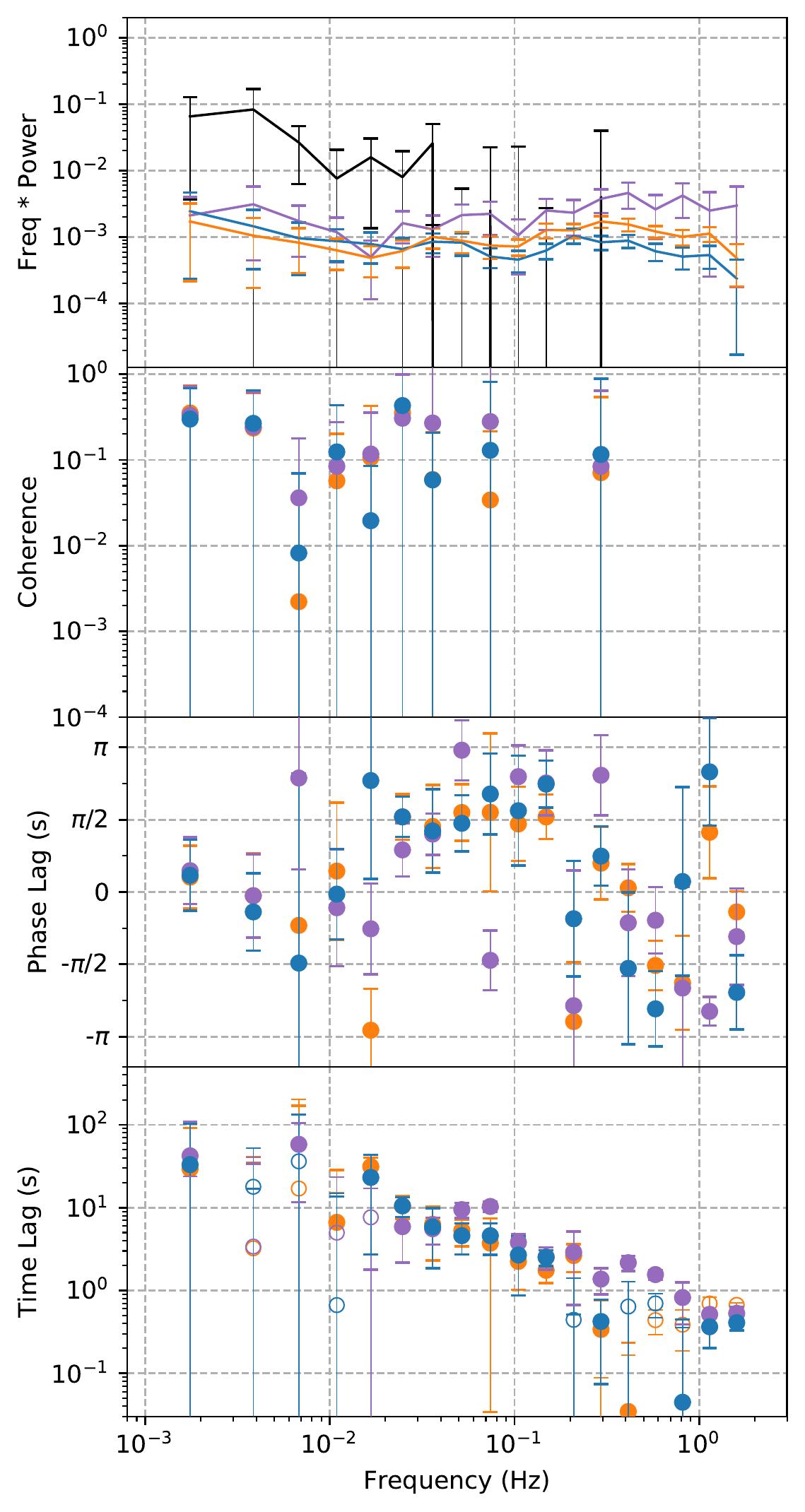} }
\hspace{4mm}\subfloat[ULTRACAM/NICER 2019 March 4]
{\includegraphics[width=0.31\textwidth]{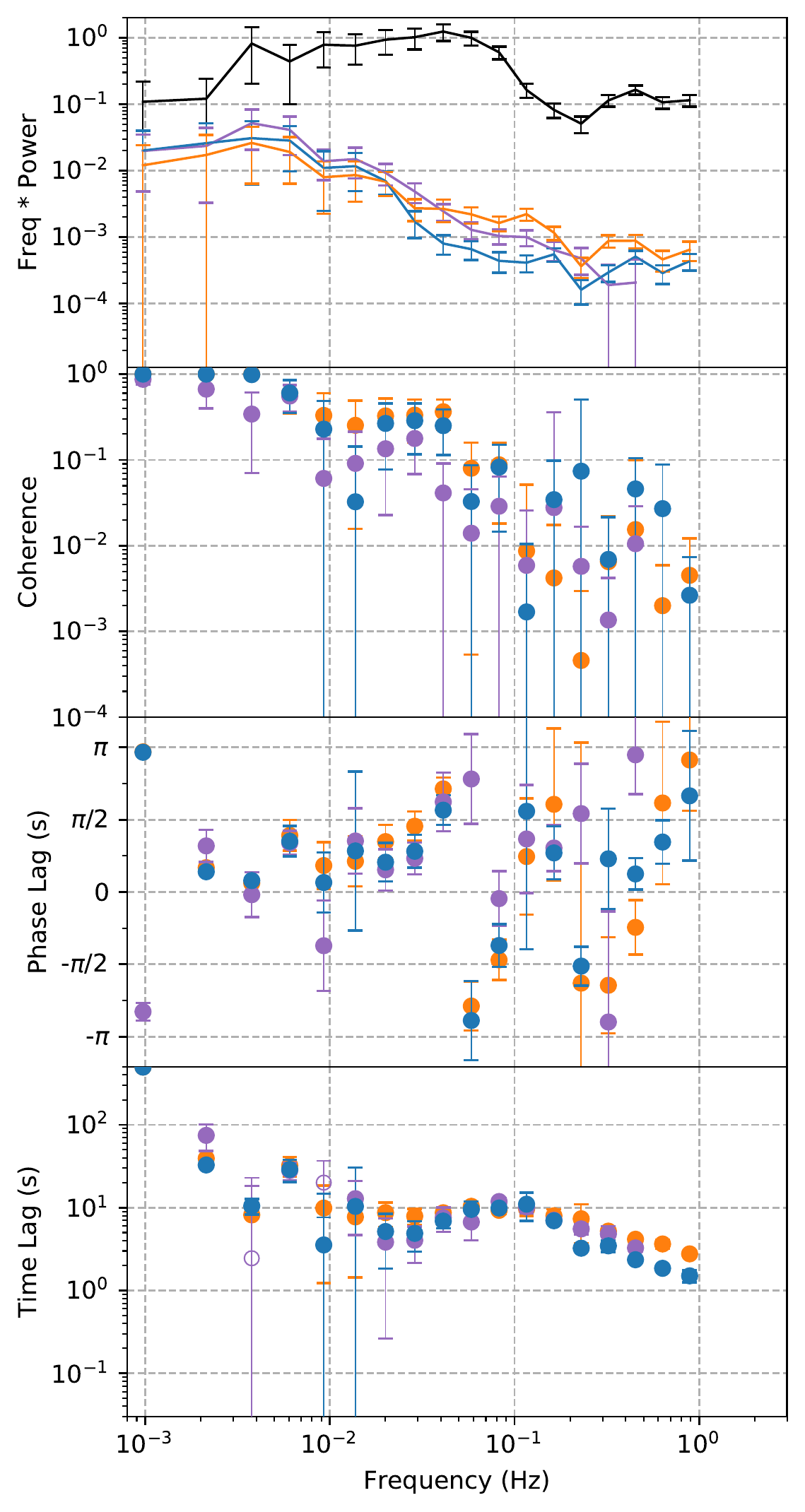} }
\hspace{4mm}\subfloat[HiPERCAM/NICER 2019 June 7]
{\includegraphics[width=0.31\textwidth]{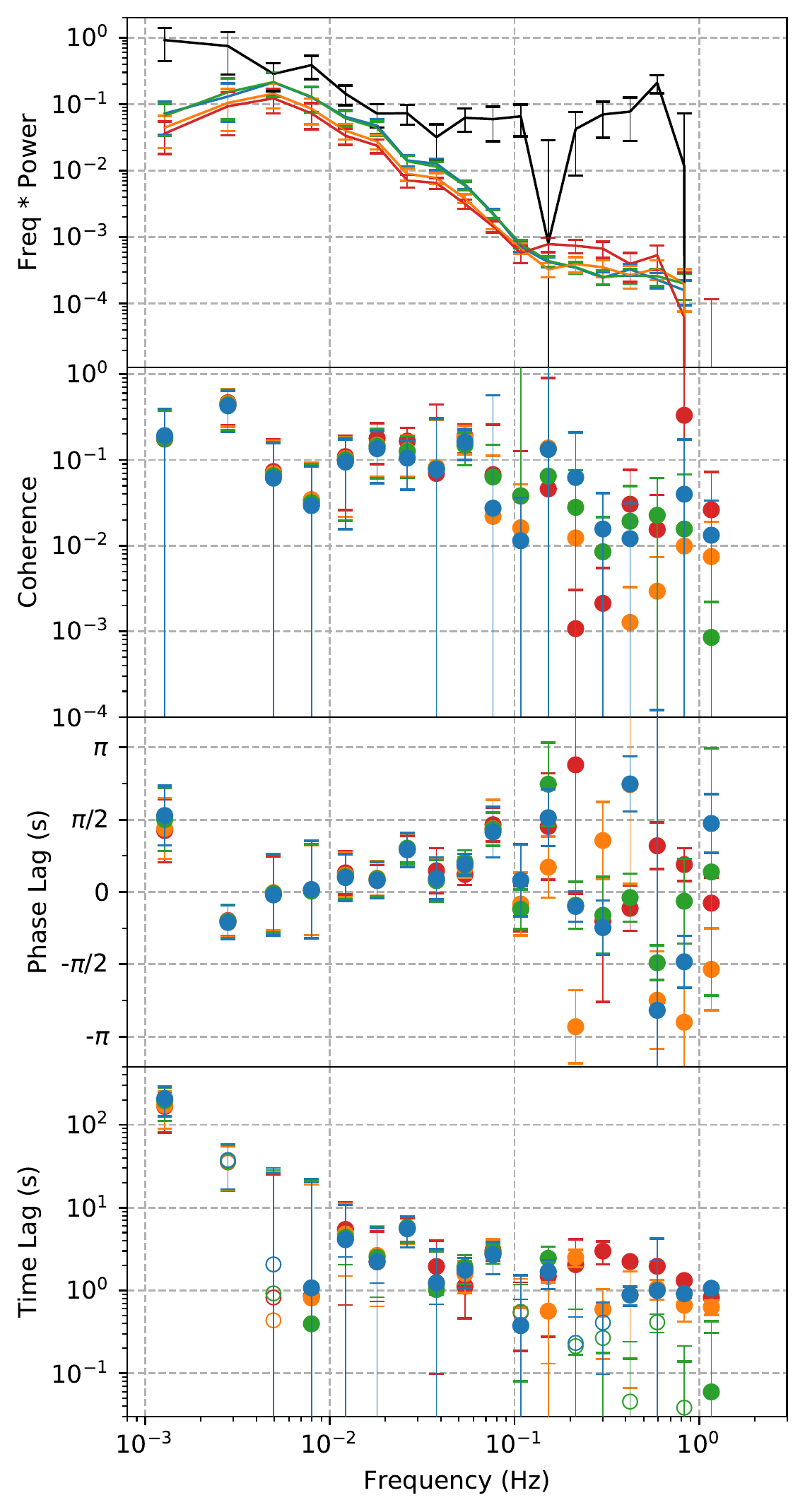} }
\caption{Results of the Fourier analysis of the simultaneous optical and X-ray light curves of \target. From top to bottom: the X-ray and optical power spectra where the white noise has been removed, coherence spectrum, phase lags and time lags. For the bottom two panels,a positive lag mean that the optical lags the X-rays. We use a logarithmic rebinning of a factor of 1.4 to display the data. In each plot, the X-ray data are shown in black, whereas the purple, blue, green, orange and red show the  $u_s$, $g_s$, $r_s$, $i_s$- and $z_s$-band data, respectively.}
\label{fig:fourier} 
\end{figure*}

\subsection{Fourier Analysis}

In order to understand the nature of the different components contributing to the CCF, we decomposed the observed variability into different time-scales using Fourier techniques. We performed a Fourier analysis of the light curves using the X-ray spectral-timing software package \textsc{stingray}\footnote{\url{https://github.com/StingraySoftware/stingray}} \citep{Huppenkothen19}. The coherence and corresponding errors were determined using the method described in \citet{Vaughan97}. We computed the Fourier transform of the light curves and then analysed them at each frequency. The power spectra represent the amplitude of the variability at each Fourier frequency, the coherence shows how the variability in the power of the correlated signal is distributed over the Fourier frequencies, and the phase lags represent a measure of the lag between the bands at each frequency as a function of phase. Sometimes, the time lags are a more intuitive representation of the delays, which are connected to the phase lags through $\Delta t = \Delta \phi/2\pi f$, where $f$ is the frequency of the bin and $\phi$ is the phase lag. Positive phase lags correspond to the delay of the optical light curve with respect to the X-rays. 

Good Time Intervals (GTIs) are used based on the individual epoch of the X-ray observations and the average cross-spectrum is computed over independent light curve segments with 2048 bins in length. We use 2, 1, and 6 segment(s) for the 2019 March 2, 2019 March 4 and 2019 June 7 data, respectively, where the white noise is fitted to each power spectrum and removed prior to the calculation of coherence. The standard root-mean-squared (RMS) normalisation is applied \citep{Belloni90b}. In Fig.\,\ref{fig:fourier} we show the frequency-dependent products binned logarithmically in frequency. The HiPERCAM data were binned by a factor of 8, and then all data were averaged over segments of 2048 bins, or $\sim$572, 1028, and 784\,s respectively (except for the $u_s$ band data in March 4, which was co-added and thus sampled differently; this was averaged over segments of 1024 bins, or 1028\,s).

For \target\ on the nights where there is significant optical and X-ray variability (2019 March 4 and June 7), the power spectra for the optical and X-ray light curves are very similar. However, there is consistently higher power in the X-ray variability compared to the optical, which suggests that the optical variability is a result of reprocessing of faster X-ray variations at frequencies above the optical power spectrum peak.

In the 2019 March 4 and June 7  data, the coherence function shows a linear decline with increasing frequency. The declining absolute value of the optical/X-ray coherence means that a single component is not a good representation of the broad-band variability. In the 2019 March 4 and June 7 data there is a plateau in the coherence between 0.01--0.1\,Hz at $\sim$10\,s (most notable in the March 4 data), during which a rise in phase lags is observed.  Beyond $\sim$0.1\,Hz, the data become white-noise dominated, and it is not possible to find meaningful results. Frequency-dependent time-lags are also observed. Below $\sim$0.01\,Hz, the time lags rise towards low Fourier frequencies and a plateau is observed between 0.01--0.1\,Hz. Beyond $\sim$0.1\,Hz the time-lags are  observed to decrease with frequency, a natural consequence of the large scatter and randomly distributed phase lags. The time lag observed at $\sim$0.1\,Hz on March 4 is longer than what is observed in the CCFs; this is likely because the lower frequency lags contribute more to the CCF than the high frequency lags, as evidenced by the higher coherence below 0.05 Hz, and the sharp drop thereafter with time lags between 3--10\,s. This combination gives rise to a culmination of many frequencies which results in the time-lag of $\sim$5\,s observed in the CCFs.

\begin{figure}
\centering
\includegraphics[width=1.0\linewidth,angle=0]{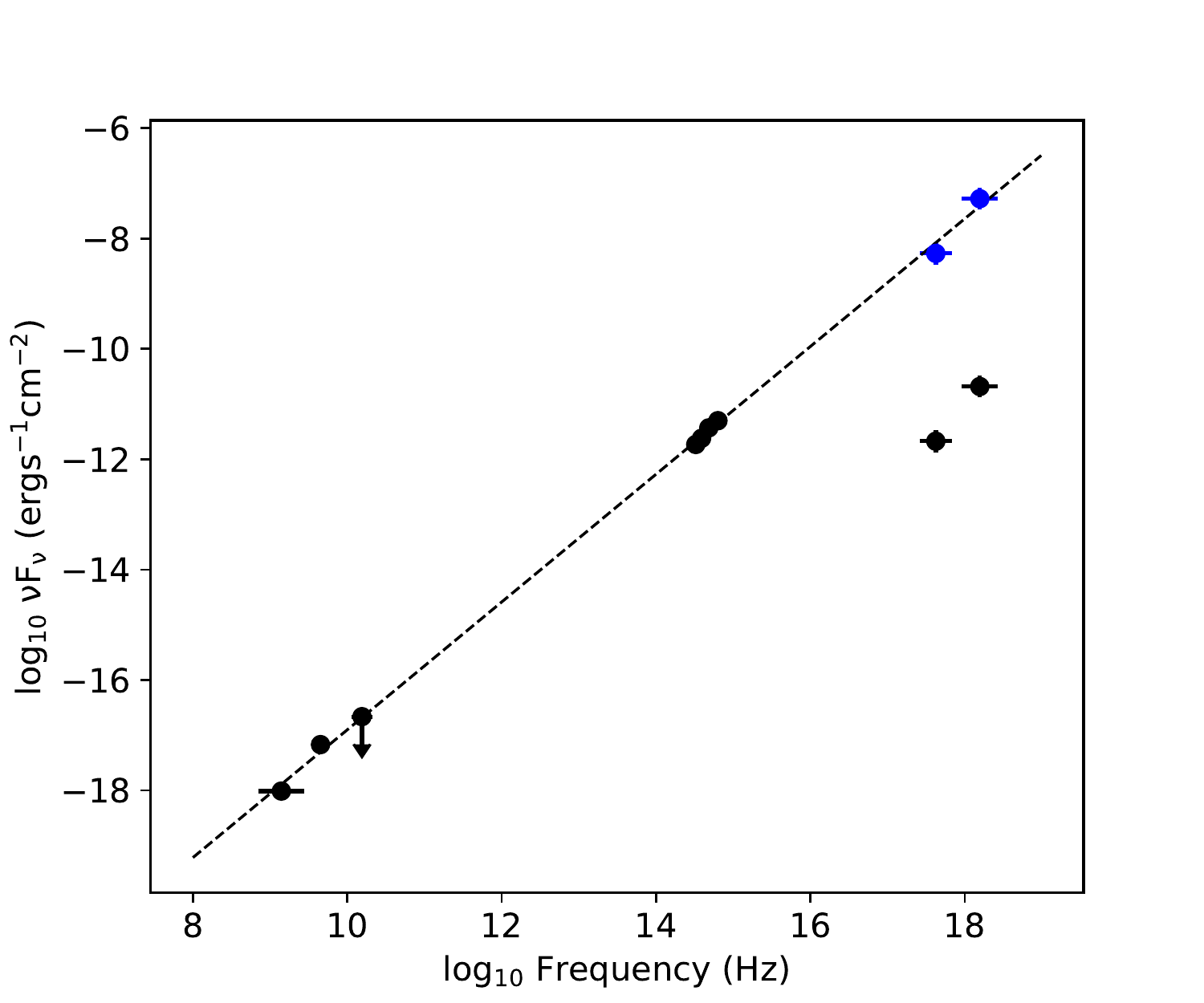}
\caption{The absorption-corrected spectral energy distribution of \target\ on 2019 June 7 (black points). The NICER (0.5--3.0\,keV and 3.0--10.0\,keV) and optical ($g_s,r_s,i_s,z_s$) data are simultaneous, whereas the radio data (1.4, 4.5, and 15.5 GHz) are interpolated values taken within $\sim$2 months \citep{Bright18, Eijnden20, Rhodes22}. We assume a distance of 12.8\,kpc. \citep{Buisson21}. The blue points are the scaled X-ray luminosities according to the $L_{\rm OIR}-L_{\rm X}$ relation of \citet{Russell06}. The radio--X-ray spectral energy distribution can be described with a power-law of the form $F_\nu \propto \nu^\alpha$ with an index of $\alpha=-0.84\pm0.02$ (dashed line).} 
\label{fig:hcam_flare_sed}
\end{figure}

\section{Discussion}

\subsection{Flare spectra}

Generally, in the optical/near-IR region, a negative power-law index is expected if there is an optically-thin synchrotron spectrum from a flow/jet, whereas a positive  power-law index is expected (with spectral index $\sim$ 1) if the optical emission is dominated by  blackbody emission from regions in the accretion disc \citep{Hynes05}. The fast, `red' optical flares observed in the ULTRACAM and HiPERCAM data have a power-law index of $\alpha \sim $-1.3, steeper than what is typically observed in XRBs, $\alpha \sim$-0.7 \citep{Hynes03,Gandhi11,Russell13}. However, it should be noted that flares with similarly steep spectral properties have been observed before with a power-law in the range -1.3 to -1.5  \citep{Russell10, Russell13, Shahbaz13, Gandhi16} and indeed the fast `red' flares are reminiscent of the `red' flares observed during the outburst of V404\,Cyg \citep{Gandhi16}. Indeed, for V404\,Cyg, based on the cooling timescales of the flaring events the emission has been attributed to synchrotron processes \citep{Dallilar17}. For optically thin synchrotron emission, the only parameter which changes the spectral index is the particle energy distribution ($p$) of the emitting electrons, which is related to the  observed spectral slope in the optically thin plasma; $\alpha_{\rm thin} = \left(1-p\right)$/$2$. If the observed quiescent power-law index of $\alpha \sim -1.3$ is interpreted as optically thin synchrotron, then $p$ = 3.6, which is steeper than $p \sim 2.4$ (or $\alpha \sim -0.7$),  which is typical for optically thin synchrotron in XRBs. A mixture of thermal and non-thermal particle energies could potentially explain such a steep slope observed in \target.  In contrast, slow, large `blue' flares are observed with a power-law index $\sim$ 1.0,  consistent with blackbody emission from an irradiated accretion disc \citep{Hynes05}; the spectrum from an irradiated accretion disc has a power-law index of 1.2 in the $g_s$ to $i_s$ bands.

We estimate the binary separation to be $\sim$5.1 \Rsun ($M_{\rm 1}$=2.0\Msun, $M_{\rm 2}$=0.25\Msun, $P_{\rm orb}$=0.883\,d)  assuming that the accretion disc extends to its tidal truncation radius ($R_{\rm d}$ = 0.9 $R_{\rm L1}$, where $R_{\rm L1}$ is the equivalent radius of the Roche lobe of a sphere with the same volume), we find $R_{\rm d} <$ 2.5$\pm$0.2\Rsun. The large flare events  can be represented by a $\sim$14,000$\pm$2,000\,K blackbody and an equivalent blackbody radius of $\sim$1.0$\pm$0.2\Rsun\ (see Section\,\ref{sec:sed}) which is consistent with arising from regions in the accretion disc or from an extended disc atmosphere or wind \citep{Buisson21}. The optical multi-wavelength spectral properties are reminiscent of those observed in the  black hole X-ray binary V404\,Cyg, where slow, `blue' as well as fast, `red' flares were observed during its 2015 outburst \citep{Kimura16, Gandhi16}. From the strongest observed ﬂare on 2019 June 6  we estimate the optical ($u_s$--$z_s$) and X-ray (0.5--10\,keV) unabsorbed flare power to be  $\sim$0.1\% and $\sim$0.33\% of the Eddington luminosity, assuming a 1.8\Msun neutron star and a distance of 12.8\,kpc \citep{Buisson20_burst}. The optical flare in \target\ is a factor $\sim$5 less powerful compared to the optical flares in GX\,339-4 \citep{Gandhi10} and V404\,Cyg  \citep{Gandhi16}.

\subsection{Spectral energy distribution}

The radio--X-ray spectral energy distribution of the X-ray binary systems GX\,339--4 \citep{Gandhi10}, MAXI\,J1820+070 \citep{Rodi21} and GRS\,1716--249 \citep{Bassi20}, can be described by a combination of non-thermal emission of electrons accelerated in the jet by internal shocks  \citep{Malzac13, Malzac14} and emission from the irradiated disc and hot corona \citep{Gierlinski09}. In Fig.\,\ref{fig:hcam_flare_sed} we show the absorption-corrected spectral energy distribution of \target\ observed on 2019 June 7 using $N_{\rm H}=1.84 \times 10^{21}$\,cm$^{-2}$. The absorption-corrected NICER soft- and hard-band  fluxes were determined  using the XSPEC software package \citep{Arnaud96} with the \textsc{tbabs(diskbb+bbody)} model with $\Gamma=1.6$.  The mean absorption-corrected  optical ($g_s,r_s,i_s,z_s$) data  were determined using the light curves  in Section\,\ref{sec:obs:hcam}. There are not many radio measurements in 2019, so we interpolate the radio flux values at 1.4, 4.5, 15.5\,GHz given in \citet{Rhodes22}, \citet{Eijnden20} and \citet{Bright18}, respectively. From the mean optical ($u_s$--$z_s$) and X-ray (0.5--10\,keV) unabsorbed fluxes on 2019 June 7 we estimate luminosities of $\sim 2.5\times 10^{35}$\erg  and $\sim 4.5\times 10^{35}$\erg, respectively, assuming a distance of 12.8\,kpc \citep{Buisson20_burst}. The optical to X-ray luminosity ratio $L_{\rm opt}/L_{\rm X}$ ratio is $\sim$0.6, which is much higher than what is typical of X-ray binaries in outburst. In neutrons star X-ray binaries the optical and X-ray luminosity's are described by $L_{\rm OIR} = 10^{10.8} L_{\rm X}^{0.63}$, where the optical luminosity is dominated by X-ray reprocessing with an additional contributions from a jet and the viscously heated accretion disc. \citep{Russell06}. We find that either the optical luminosity in \target\ is a factor of $\sim$140 more than what is expected or that X-ray luminosity is a factor of $\sim$2530 under-luminous.  Note that the 2019 June 7 observations were taken at orbital phase $\sim$0.93 and given the high binary inclination angle \citep{Buisson21,Knight22} the low X-ray luminosity can be explained by optically thick material in the outer regions of the accretion disc or secondary star blocking most of the direct X-ray emission. So what we observe is scattered X-rays and the intrinsic X-rays is much higher. If we scale the X-rays using the $L_{\rm OIR}-L_{\rm X}^{0.63}$ relation \citep{Russell06},  we find that the radio--X-ray spectral energy distribution can be described with a power-law of the form $F_\nu \propto \nu^\alpha$ with an index of $\alpha \sim$ 0.16.  Indeed, this is similar to what is observed in the mean spectrum of GX\,339-4 \citep{Gandhi10} and XTE\,J1118+480 \citep{Hynes03}, where the spectral energy distribution is attributed to a mixture of optically thin synchrotron emission from a jet and the irradiated accretion disc/corona.

\begin{figure*}
\subfloat[ULTRACAM/NICER 2019 March 2]
{\includegraphics[width=0.45\textwidth]{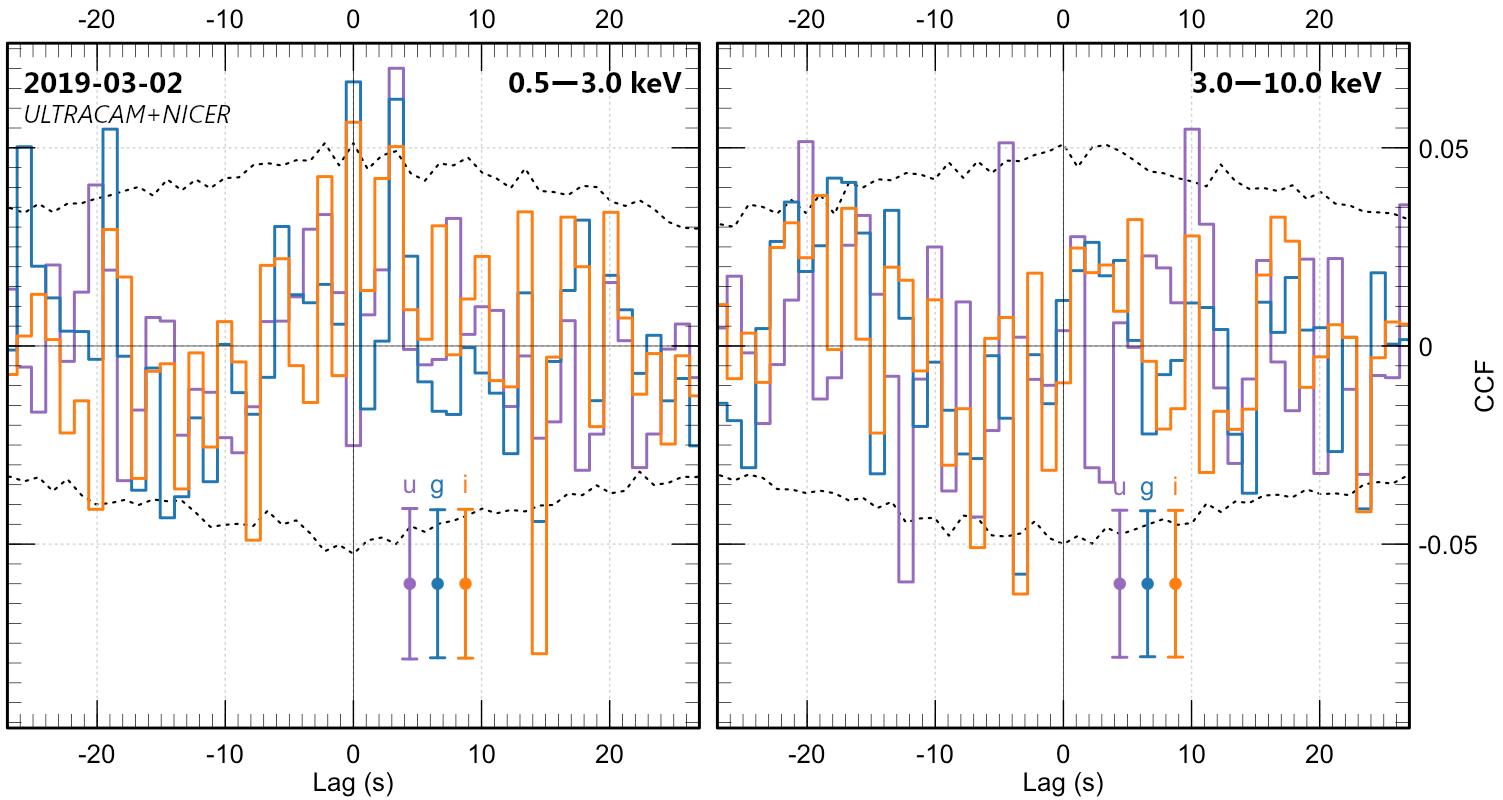}}
\subfloat[ULTRACAM/NICER 2019 March 4]
{\includegraphics[width=0.45\textwidth]{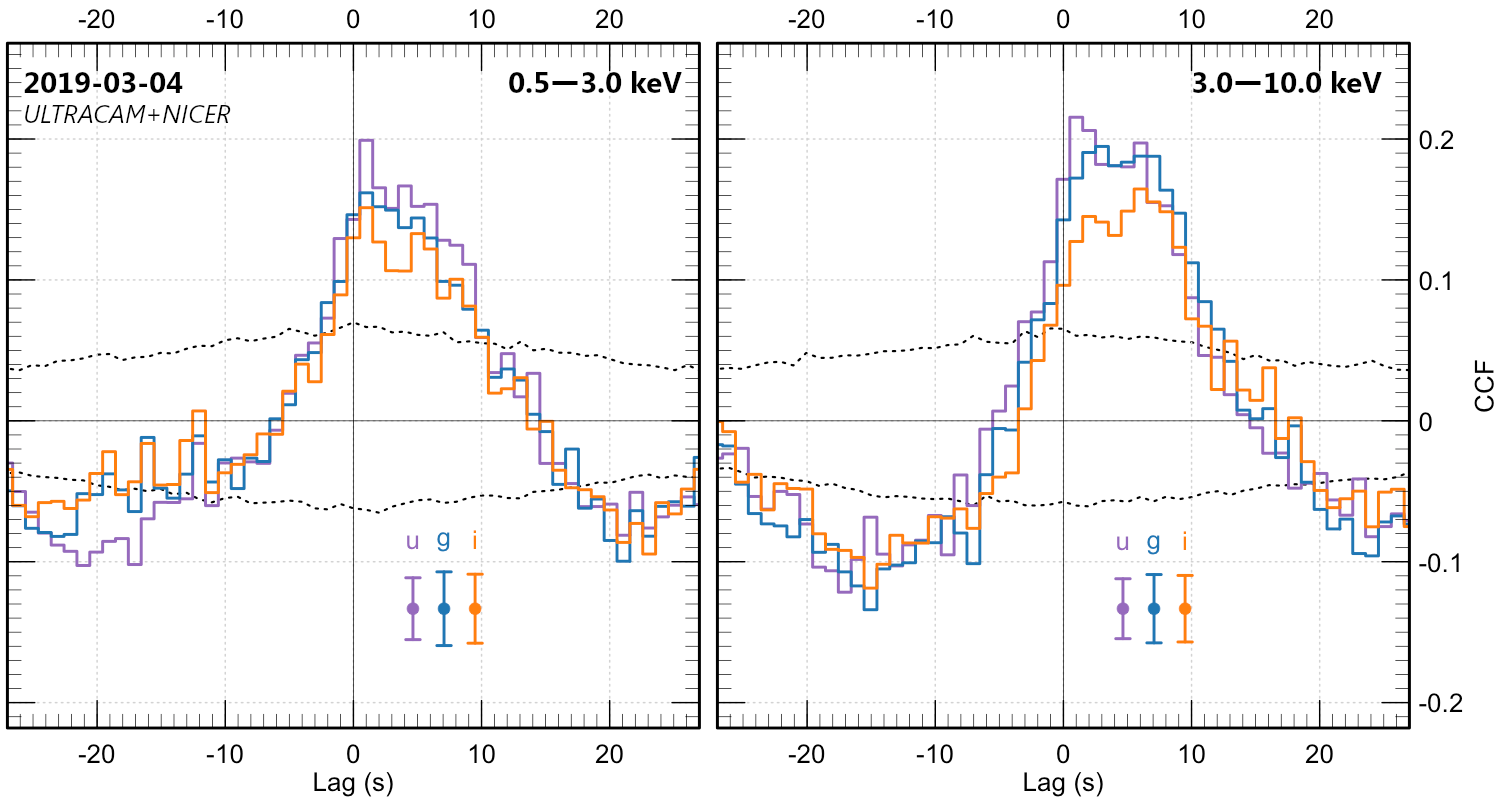}}

\subfloat[HiPERCAM/NICER 2019 June 7; part 1]
{\includegraphics[width=0.45\textwidth]{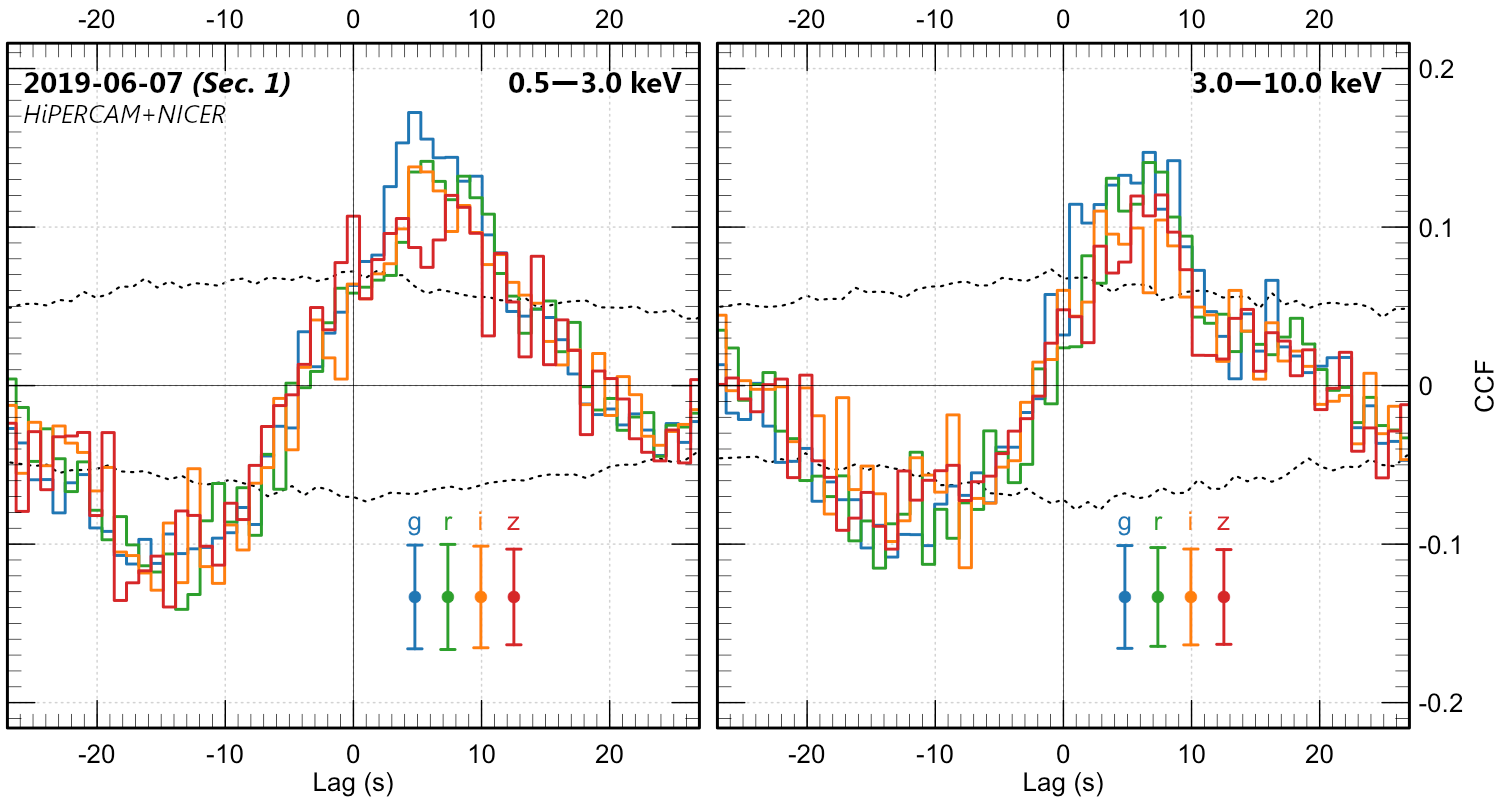}}
\subfloat[HiPERCAM/NICER 2019 June 7; part 2]
{\includegraphics[width=0.45\textwidth]{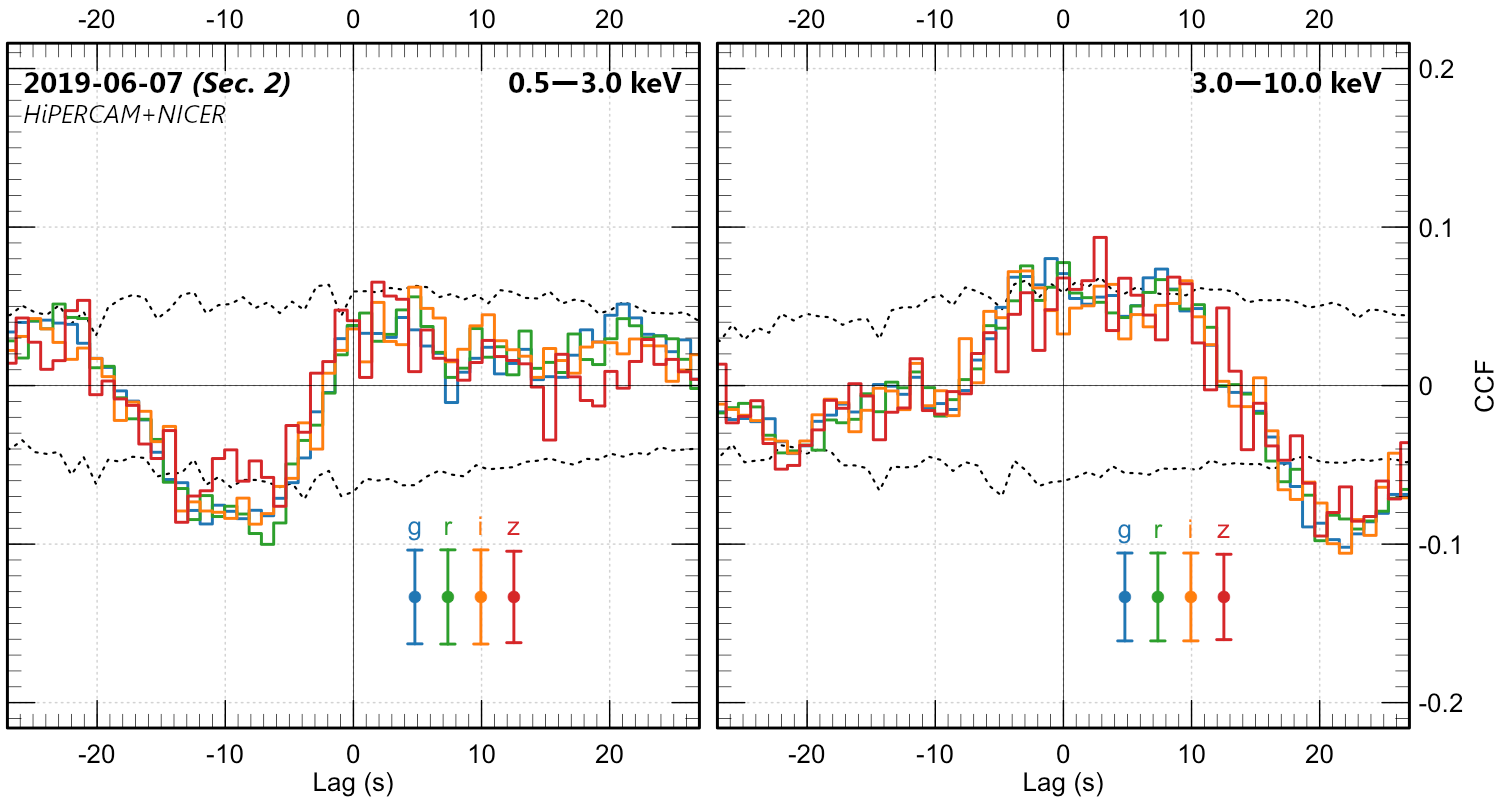}}

\subfloat[HiPERCAM/NICER 2019 June 7; part 3]
{\includegraphics[width=0.45\textwidth]{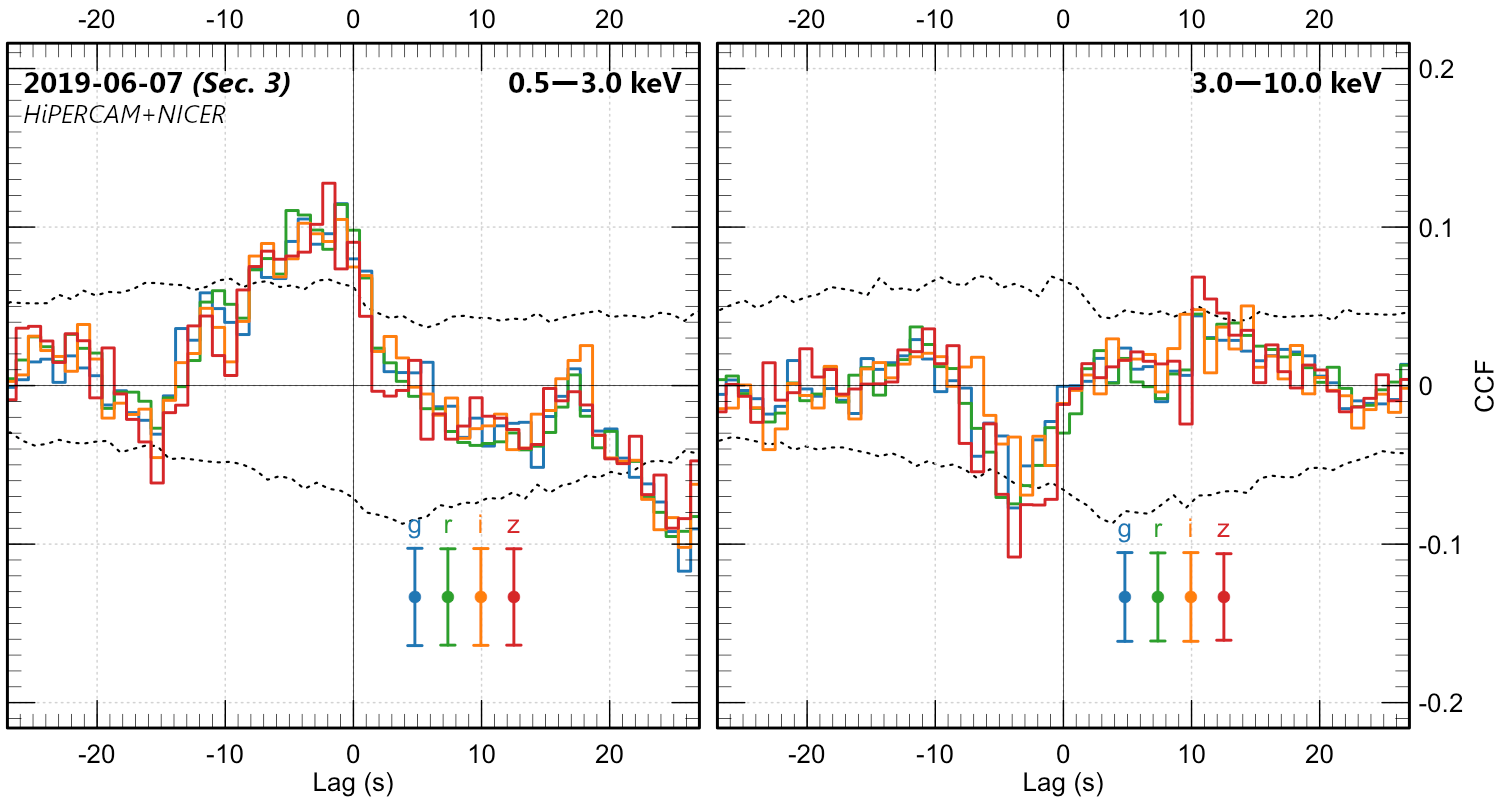}}
\subfloat[HiPERCAM/NICER 2019 June 7; part 4]
{\includegraphics[width=0.45\textwidth]{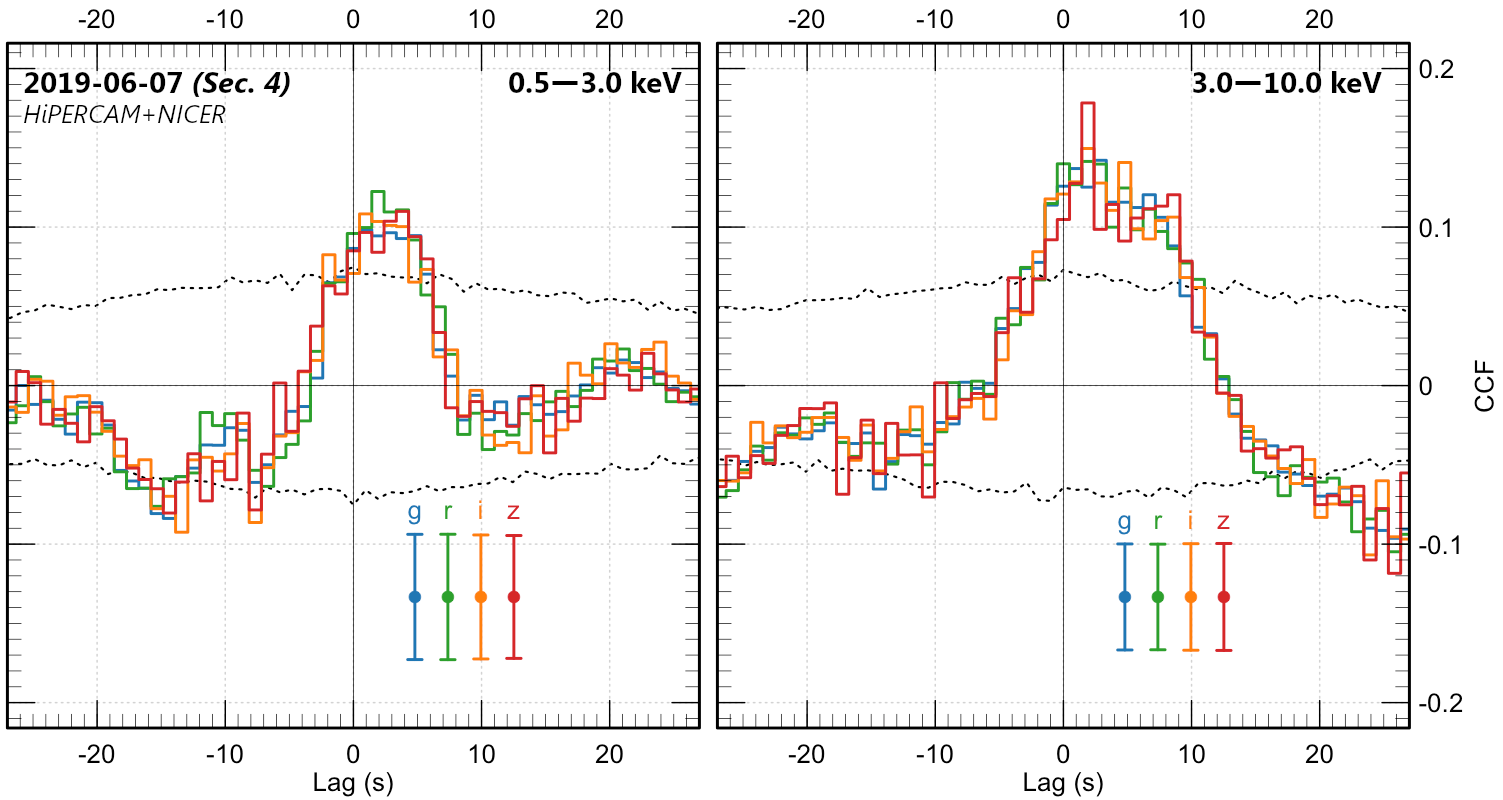}}

\caption{The HiPERCAM/NICER CCFs. The left plot shows the CCF of the optical bands versus soft X-rays (0.5--3.0\,keV) and the right plot shows the CCF of the optical bands versus hard X-rays (3.0--10.0\,keV). The black dashed lines represent the 5 and 95 percent confidence intervals.}
\label{fig:ccf_energy} 
\end{figure*}

\subsection{Optical/X-ray correlations}

In the optical waveband,  many components can potentially contribute to the optical emission e.g. the irradiated secondary star, the cold optically-thick accretion disc, the hot optically-thin X-ray emitting medium and hot flow/jet \citep{Poutanen14}. Whereas in the X-rays, two separate components are present, a soft component arising from Comptonization of disc photons and a harder component arising from synchrotron Comptonization in the hot flow \citep{Veledina16}. Indeed, this results in optical/X-ray correlations that show complex patterns, with both positive and negative correlations. The CCFs show a variety of shapes: some show positive correlations with optical photons lagging the X-rays, consistent with simple reprocessing \citep{O'Brien02, Hynes09,Paice18, Kajava19}; some show a very broad and nearly symmetric positive cross-correlation \citep{Casella10}; some show a more complex structure containing a narrow `precognition' dip at negative lags (optical photons leading X-rays) superimposed on a very broad positive cross-correlation \citep{Kanbach01, Gandhi08, Durant08, Durant11, Lasso-Cabrera13}; and some show only a  strong broad anti-correlation \citep{Motch83,Pahari17} or a narrow positive correlation superimposed on  a very broad positive cross-correlation \citep{Hynes19}. Cyclo-synchrotron optical photons undergoing Compton upscattering to X-rays in a hot flow can also reproduce both the observed optical/X-ray anti-correlation and QPOs \citep{Veledina11, Veledina13, Veledina15}.  In some cases the observed features can be explained by synchrotron emission from internal shocks within a relativistic compact jet \citep{Malzac13, Hynes19, Paice19}.  Finally, in some sources a fast optical delay component at $\sim$\,100\,ms is observed which is associated with the base of the optically--emitting  jet close to the compact object \citep{Gandhi08,Gandhi17,Paice19}.

Although there are some strong similarities in the timing behaviour of \target\ with well-studied XRBs, one notable difference is the lack of a $\sim$\,100\,ms positive optical time lag with respect to X-rays. This feature has been seen in the cross-correlated timing behaviour of three sources now: GX\,339--4 \citep{Gandhi08}, V404\,Cyg \citep{Gandhi17} and MAXI\,J1820--070 \citep{Paice19}. Both timing and multi-wavelength spectral properties support an origin of this feature in the inner jets of hard state binaries, in a compact region no larger than a few thousand Schwarzschild Radii. \citet{Malzac14} has shown that flicker noise Lorentz factor plasma variations within a compact jet can naturally produce such timing lags. The fact that \target\ does {\em not} show this feature then implies some difference between its internal jet structure with respect to other systems. Whether this is related to a difference in jet plasma Lorentz factors during the state when it was observed, or perhaps even a difference in compact object types (all three systems named above host black holes whereas \target\ does not), remains to be investigated.

In the standard reprocessing model, X-rays arising from the inner accretion disc photoionize and heat the surrounding regions, which later recombine and cool producing lower energy (optical/near-IR) photons.  The observed optical/near-IR flux is thus delayed relative to the X-rays  due to the light travel time between the X-ray source and the reprocessing region. The corresponding CCF arising from X-ray reprocessing has a characteristic orbital phase-dependent shape, where the CCF rises from negative lags, peaks, and subsequently falls off \citep{Hynes98, O'Brien02}.  Depending on the orbital phase the CCF can be very symmetric, but sometimes an extended positive delay is observed, especially near quadrature \citep{O'Brien02, Hynes09}. The shapes of the CCFs observed in \target\ are more consistent with the shape of the CCFs in Sco\,X--1, Cyg\,X--2 \citep{Durant11}, rather than other XRBs such as XTE\,J1118+480 \citep{Kanbach01}, Swift\,J1753.5--0127 \citep{Durant08} and GX\,339--4 \citep{Gandhi08} and MAXI\,J1820+70 \citep{Paice19},  where `pre-recognition' dips are observed and X-ray reprocessing is not thought to be dominant. The time delay between the optical/near-IR and X-ray flux can be up to twice the binary separation ($a$) and can be obtained from Kepler's third law: $a/c = 9.77\,M^{1/3} P_{\rm d}^{2/3}$\,s  (where $c$ is the speed of light, $M$ is the sum of the binary masses in solar units and $P_{\rm d}$ is the orbital period in days). Although the binary parameters for \target\ are not fully known, the orbital period of 21.2\,hr together with estimates of the binary masses allows one to estimate the binary separation to be $a/c \sim$ 12\,s. Indeed, we observe CCFs with time delays of $\sim$5--15\,s which suggests that the delays are consistent with arising from regions in the accretion disc.
 
As mentioned earlier, in the hybrid hot inner flow model of \citet{Veledina16} two X-ray components, one arising from disc Comptonization and the other from synchrotron Comptonization, as well as two optical components due to  synchrotron self-Compton emission from the hot inner accretion flow and disc reprocessing are present. In the X-rays, the seed photons for Comptonization are provided by the accretion disc (disc Comptonization) which dominates in the hard state. However, the hot flow itself also produces synchrotron radiation that can contribute or even dominate the seed photon flux for Comptonization (synchrotron Comptonization). In the optical, the flux can arise from X-ray reprocessing or from synchrotron emission in the hot inner accretion flow. An anti-correlation and negative lags between the optical and X-ray flux is expected because the increase in the mass accretion rate leads to an increased X-ray flux and a higher level of synchrotron self-absorption,  leading to a drop in the optical emission \citep{Veledina11}. Furthermore, the  optical is expected to have a stronger anti-correlation with the hard X-rays compared to with the soft X-rays,  characteristics that are expected if the source transitions from a hard to soft state. During the initial stages of the outburst of \target\ (in the hard state) we observe CCFs with a positive peak at a time delay of $\sim$5--15\,s and optical ACFs which are broader than the X-ray ACFs (see Fig.\,\textcolor{blue}{\ref{fig:acf_ccf}a,b}). This implies some underlying connection between the optical and X-ray fluxes and is consistent with optical flux arising from X-ray reprocessing in the outer regions of the accretion disc. For example, the 2019 March 4 CCF shows a nearly symmetric positive correlations at positive lags which is consistent with X-ray reprocessing, supported by the wavelength dependant optical/X-ray delays in the CCFs, in which the longest  wavelength delay has the longest delay. On the other hand, the 2019 June 7 data taken during a softer state cannot be described within the simple reprocessing scenario. The narrow optical ACF (comparable with the X-ray ACF) and the negative correlation in the optical/X-ray CCFs (see Fig.\,\textcolor{blue}{\ref{fig:acf_ccf}c--f}) are the characteristics of the synchrotron self-Compton mechanism operating in a hot accretion flow \citep{Veledina11}. The presence of both synchrotron and reprocessed X-ray emission in the optical is in line with the spectral energy distributions of the observed fast `red' flares (see Section\,\ref{sec:sed}). 

The CCFs of the 2019 June 7 parts 1 and 4 data have similar shapes, with anti-correlations at negative lags and positive correlation at positive lags. The shape can be explained by the presence of two emission components in the optical, with the X-rays being dominated by the synchrotron Comptonization continuum \citep{Veledina17}. The CCF of the 2019 June 7 part 3 data shows a hint of positive correlation at negative lags. It looks very similar to the CCF observed in MAXI\,J1820+070 \citep[see epoch 6 in][]{Paice21}. To explain this shape, one requires an additional source of X-ray photons arising from the disc Comptonization. Indeed, the hard-to-soft spectral state transition involves the motion of the cold accretion disc towards the compact object. As the role of the disc increases with the overall increase in the mass accretion rate, the power dissipated in the hot accretion flow increases, so the whole spectrum of this component increases (similar to ADAFs) resulting in the enhancement of the synchrotron emission. The simultaneous presence of two X-ray components, synchrotron Comptonization and disc Comptonization, leads to the complex shape of the optical/X-ray CCF and manifests itself through the different correlations with the soft and hard X-ray bands.

To investigate this possibility further, we separate the X-ray range into soft (0.5--3.0\,keV) and hard (3.0--10.0\,keV) energy bands and show the CCFs with respect to only one optical band ($g_s$) for clarity (see Fig.\,\ref{fig:ccf_energy}). We systematically observe different correlations between the optical and soft/hard-X-rays, supporting the assumption of two X-ray components. In Appendix\,\ref{appendix:model} we attempt to reproduce the timing and correlation properties observed in \target\ in the context of the hot inner flow-disc Comptonization and reprocessing model \citep{Veledina11,Veledina18}. The low absolute value of the optical/X-ray coherence of $\sim$0.1--0.2 means that multiple components are required to explain the broad-band variability. We clearly observe correlations between some optical and X-ray flares which shows that they are indeed related,  some flares events have weak correlations and so may not be related. In general, we find good qualitative agreement between the data and the multi-component hot inner flow-disc Comptonization and reprocessing model, and find that the relative role of the different X-ray and optical components vary during the course of the outburst as well as on shorter time-scales. 

\section{Conclusions}

We present a rapid timing analysis of simultaneous optical (HiPERCAM and ULTRACAM) and X-ray (NICER) observations of the X-ray transient \target\ during  2018 and 2019. The optical light curves show rapid, small amplitude ($\sim$0.1\,mag in $g_s$) `red' flares (i.e. stronger at longer wavelengths) on time-scales of $\sim$seconds which have a power-law index consistent with optically thin synchrotron emission. The optical light curves also show relatively slow, large amplitude ($\sim$1\,mag in $g_s$) `blue' flares (i.e. stronger at shorter wavelengths) on time-scales of $\sim$minutes, with a spectral energy distribution consistent with X-ray reprocessing in the accretion disc. 

We present a Fourier time- and energy-dependant timing analysis of the simultaneous optical/X-ray light curves. The simultaneous optical and X-ray data show correlated variability  that has a strong hard-energy component on 2019 March 2 and 4, and a strong soft-energy X-ray component on 2019 June 7, suggesting a spectral state change. We find that the optical ACF is broader than the X-ray ACF during the initial outburst stages, which can be explained by simple X-ray reprocessing.  The coherence function shows a linear decline with increasing frequency. There is also a plateau in the time lags between 0.01--0.1\,Hz at $\sim$10\,s. These characteristics can be attributed to thermal reprocessing of X-ray emission in the outer regions of the accretion disc.

We find that relative roles of the different X-ray and optical components governs the shape of the optical/X-ray CCFs and vary on shorter time-scales. The CCFs of the simultaneous optical versus soft- and hard-band X-ray light curves show time- and energy dependent correlations. The 2019 March 4 and 2019 June parts 1 and 4 CCFs show a nearly symmetric positive correlations at positive lags consistent with simple X-ray disc reprocessing. The soft- and hard-band CCFs  are similar and can be reproduced if disc reprocessing dominates in the optical and one component (disc or synchrotron Comptonization) dominates both the soft and hard X-rays. The 2019 June 7 part 3 data obtained between parts 1 and 4, shows a very different CCFs. The observed positive correlation at negative lags in the soft X-ray band can be reproduced  if the  optical synchrotron emission is correlated with the hot flow X-ray emission. The observed timing properties are in qualitative agreement with the inner hot accretion flow model, where X-rays are produced by both synchrotron and disc Comptonization and the optical emission arises from the hot flow synchrotron and irradiated disc components. 

\section*{Acknowledgements}

TS and VSD acknowledge financial support from the Spanish Ministry of Science, Innovation and Universities (MICIU) under grant PID2020-114822GB-I00.
KMR acknowledges funding from the European Research Council (ERC) under the European Union's Horizon 2020 research and innovation programme (grant agreement No 694745).
PG and JAP acknowledges support from Science and Technology Facilities Council (STFC) and a UGC-UKIERI Thematic Partnership.
TRM acknowledges support from STFC, grant ST/T000406/1.
M.R.K acknowledges support from the Irish Research Council in the form of a Government of Ireland Postdoctoral Fellowship (GOIPD/2021/670: Invisible Monsters). 
M.R.K., R.P.B., and C.J.C. acknowledge support from the ERC under the European Union’s Horizon 2020 research and innovation programme (grant agreement No.715051; Spiders).
The design and construction of  HiPERCAM was funded by the European Research Council under the European Union's Seventh Framework Programme (FP/2007-2013) under ERC-2013-ADG Grant Agreement no. 340040 (HiPERCAM).
HiPERCAM operations and VSD are supported by STFC grant ST/V000853/1.

Based on observations were made with the Gran Telescopio Canarias, installed at the Spanish Observatorio del Roque de los Muchachos of the Instituto de Astrofísica de Canarias, on the island of La Palma.
Based on observations made with ESO Telescopes at the La Silla Paranal Observatory under ESO programme 096.D-0808. 
We gratefully acknowledge the use of \textsc{python} packages: \textsc{matplotlib} \citep{Hunter07} and \textsc{numpy} \citep{vanderWalt11}.
We acknowledge to use of Aladin \citep{Bonnarel00}.
This research has made use of data and/or software provided by the High Energy Astrophysics Science Archive Research Center (HEASARC), which is a service of the Astrophysics Science Division at NASA/GSFC and the High Energy Astrophysics Division of the Smithsonian Astrophysical Observatory. 
The Pan-STARRS1 Surveys (PS1) have been made possible through contributions of the Institute for Astronomy, the University of Hawaii, the Pan-STARRS Project Office, the Max-Planck Society and its participating institutes, the Max Planck Institute for Astronomy, Heidelberg and the Max Planck Institute for Extraterrestrial Physics, Garching, The Johns Hopkins University, Durham University, the University of Edinburgh, Queen's University Belfast, the Harvard-Smithsonian Center for Astrophysics, the Las Cumbres Observatory Global Telescope Network Incorporated, the National Central University of Taiwan, the Space Telescope Science Institute, the National Aeronautics and Space Administration under Grant No. NNX08AR22G issued through the Planetary Science Division of the NASA Science Mission Directorate, the National Science Foundation under Grant No. AST-1238877, the University of Maryland, and Eotvos Lorand University (ELTE).

\medskip
\noindent
{\it Facilities:} GTC (HiPERCAM), NTT (ULTRACAM), NICER

\section*{Data availability}

The ULTRACAM and HiPERCAM data can be obtained by contacting the ULTRACAM team. The NICER data are available in the HEASARC Data Archive (https://heasarc.gsfc.nasa.gov/docs/archive.html). The data used in this paper will be shared on reasonable request to the corresponding author.

\bibliographystyle{mnras}


\appendix
\appendixpage
\addappheadtotoc

\section{Optical light curves properties}

\begin{figure*}
\centering
\includegraphics[width=1.27\textwidth,angle=90]{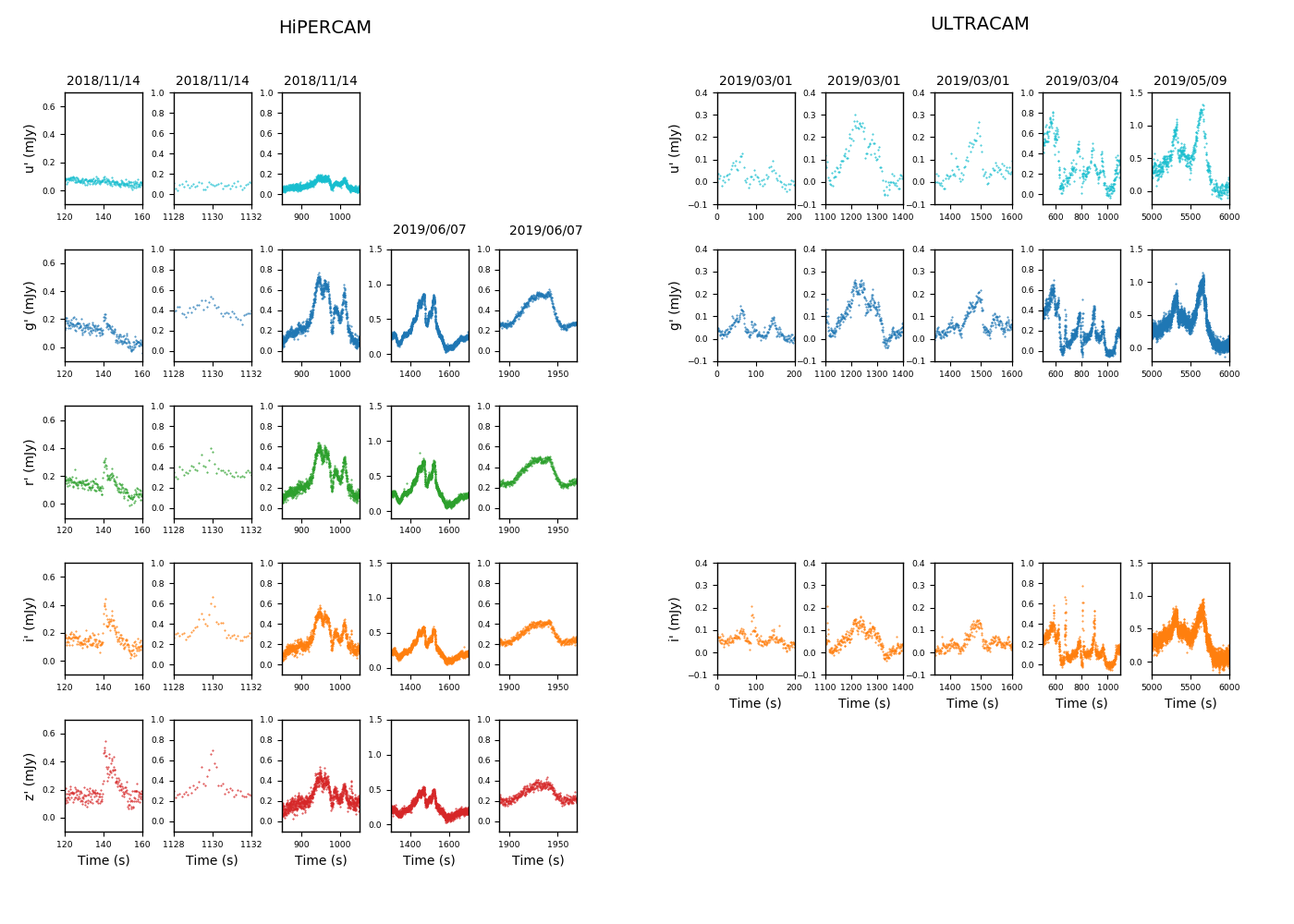}
\caption{Examples of the HiPERCAM (left) and ULTRACAM (right) small and large flare events.} 
\label{fig:flares_examples}
\end{figure*}

\begin{figure*}
\centering
\includegraphics[width=0.23\linewidth,angle=0]{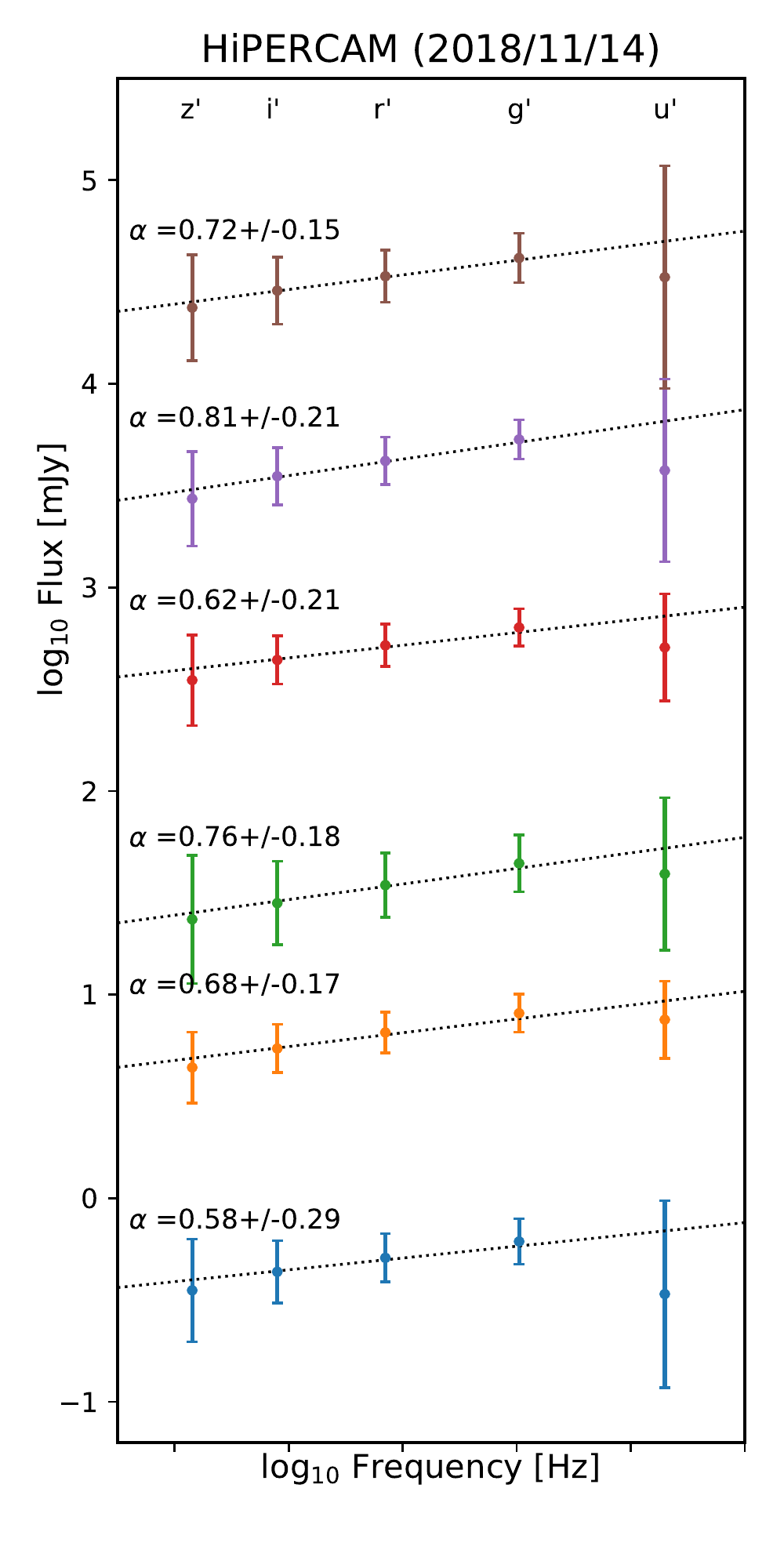}
\includegraphics[width=0.23\linewidth,angle=0]{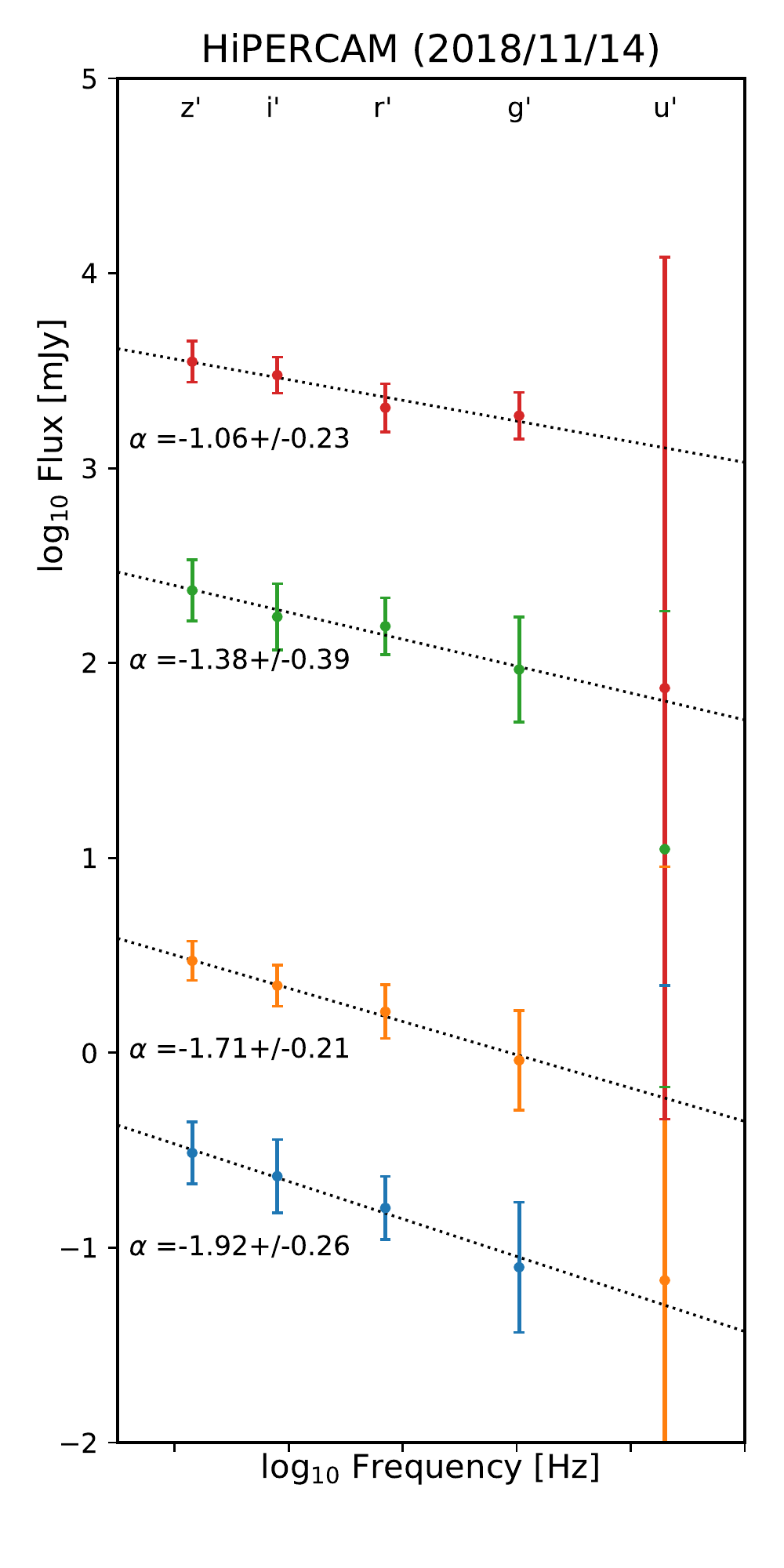}
\includegraphics[width=0.46\linewidth,angle=0]{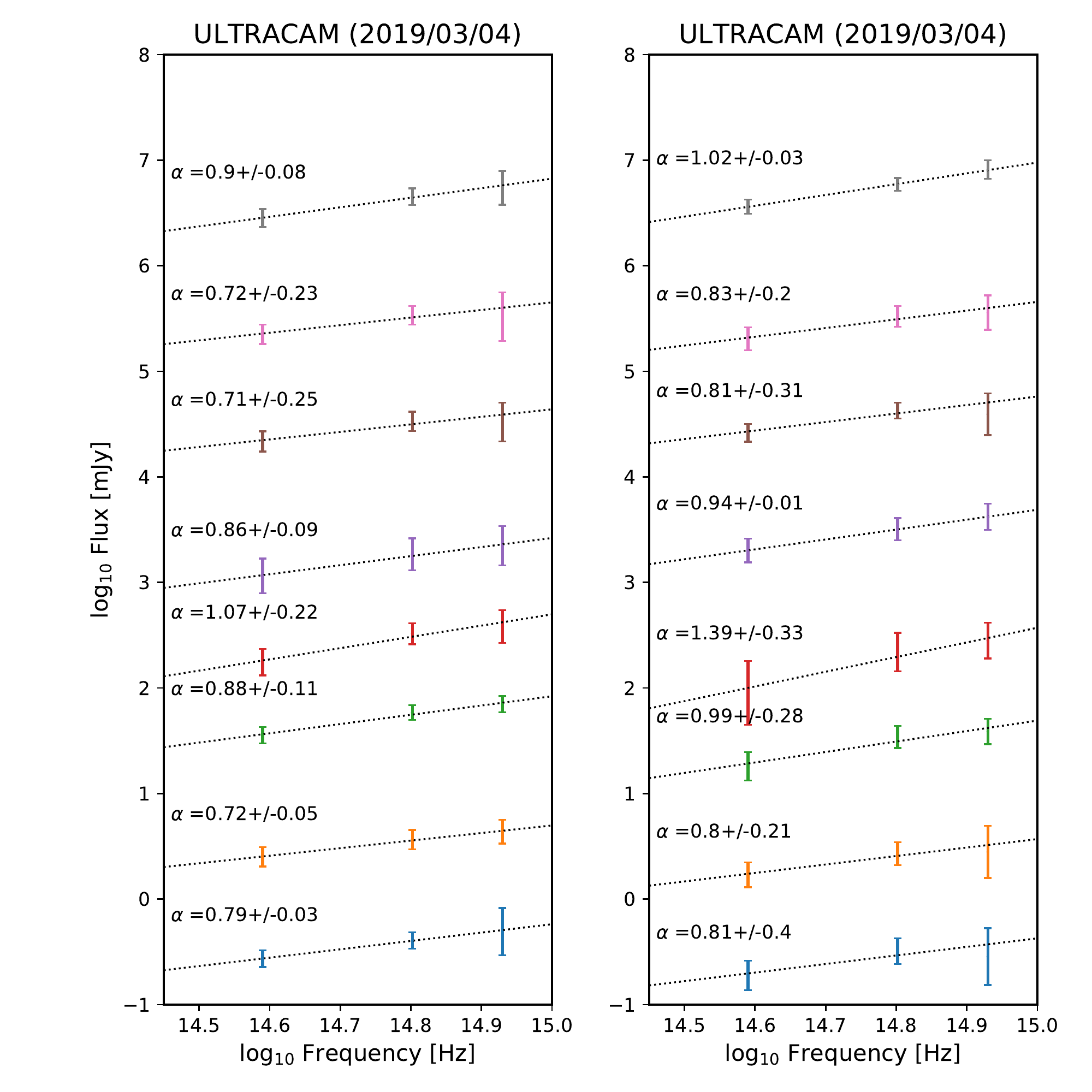}
\caption{Examples of the broad-band dereddened spectral energy distribution of the flaring events observed in \target. The dotted lines show a power-law fit to the data of the form $F_\nu \propto \nu^\alpha$. The left panel shows the HiPERCAM data taken on 2018 November 14 and the right panel shows the ULTRACAM data taken 2019 May 9.} 
\label{fig:flares_sed}
\end{figure*}

In Fig.\,\ref{fig:flares_examples} we show examples of the HiPERCAM and ULTRACAM small and large flare events. As one can see, flares on different time-scale, amplitudes and colour are present. We also show examples of fits to the broad-band dereddened spectral energy distribution of the observed flaring events  (see Fig.\,\ref{fig:flares_sed}).

\section{Modelling the timing properties in the context of the hybrid hot inner flow model}
\label{appendix:model}

The observed complex shape of the CCFs can be explained via an interplay between different optical emission components, namely, reprocessing in the accretion disc, synchrotron emission from the jet and emission from the hot inner accretion flow \citep[see e.g.][]{Veledina13, Malzac18, Paice21}. The systematically different correlations of the optical/hard X-rays versus optical/soft X-rays indicate the presence of at least two sources of X-ray photons, one dominating at soft energies and the other dominating at hard energies. Indeed, studies of the broad-band spectral and timing properties of many black hole X-ray binaries reveal several emission components in the X-rays. A standard and irradiated accretion disc dominates the soft X-rays, as demonstrated by the covariance spectra and the soft X-ray time lags \citep[see e.g.][]{Uttley11, Cassatella12, deMarco15, deMarco21};  a disc Comptonization and/or synchrotron Comptonization components stratified in a hot medium contribute to the hard X-rays, as shown by the hard X-ray time lags \citep[see e.g.][]{Kotov01, Veledina13, Mahmoud18}, as well as a complex reflection feature appearing at hard X-ray energies, studied via frequency-resolved spectroscopy and reverberation lags \citep{Revnivtsev99, Fabian10, Uttley14, Axelsson21}. When the different components contribute  approximately equally, the interplay between these components is observed  in terms of complex shapes in the X-ray power spectra and the cross-correlation function between the optical and X-ray bands  \citep{Veledina16, Veledina18}.

\begin{figure}
\includegraphics[width=0.50\textwidth]{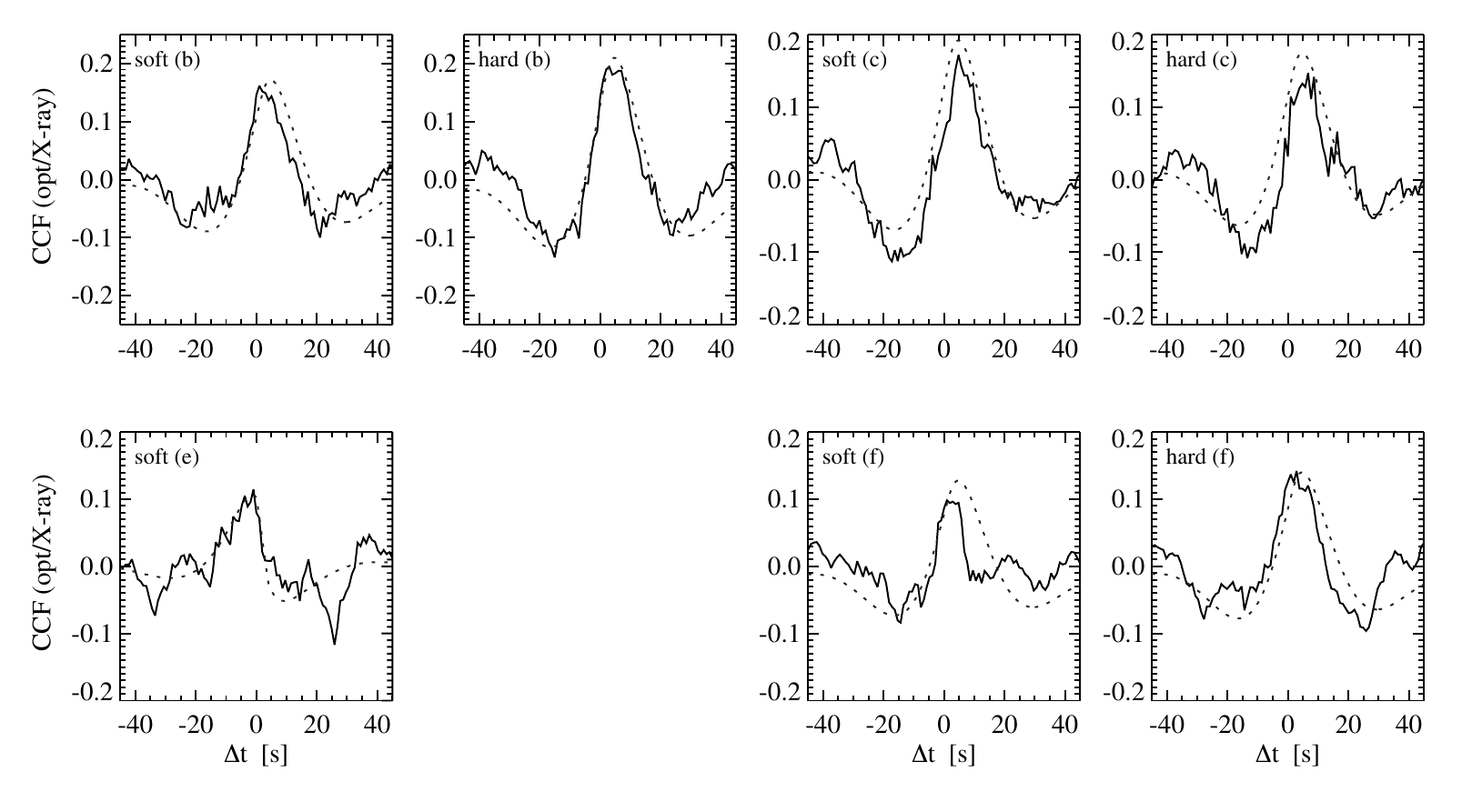}
\caption{We show the ULTRACAM or HiPERCAM optical versus NICER X-ray CCFs. The optical ($g_s$-band) versus soft (0.5--3.0\,keV) and hard X-rays (3.0--10.0\,keV) are shown. We only show the data where the CCFs are significantly above the noise level. The data taken on 2019 March 4 (b), 2019 June 7 parts 1 (c), 3 (e) and 4 (f) are shown. The solid line shows the data whereas the dashed line shows the hybrid hot inner flow model.}
\label{fig:model} 
\end{figure}

To reproduce the shapes of the observed CCFs in \target, we investigate the scenario of the simultaneous presence of two X-ray components, one arising from disc Comptonization and the other from synchrotron Comptonization, as well as two optical components due to  synchrotron emission from the hot inner accretion flow and disc reprocessing. The variability of both components are caused by accretion rate fluctuations \citep{Lyubarskii97} and the power spectra are modelled through  zero-centered Lorentzian functions. The accretion disc is assumed to be truncated at some radius away from the compact object. The synchrotron Comptonization continuum is assumed to be delayed with respect to the disc Comptonization by the time it takes the fluctuations to propagate from the radius of truncated disc to the place where the synchrotron is effectively Comptonized. The spectral shape of the synchrotron and disc Comptonization emission change under the changing mass accretion rate with two major patterns: (i) an increase and decrease in flux with a constant spectral slope  and (ii) spectral pivoting. Depending on the relative amplitudes of these variations, the number of photons in a given X-ray band may correlate or anti-correlate with the mass accretion rate fluctuations. This can be parameterised through the simple relation 

\begin{eqnarray}
 x_{\rm h}(t) &=& \varepsilon_{\rm h} \dot{m}(t+t_0) \ast g(t) + \dot{m}(t) \\
 x_{\rm s}(t) &=& \varepsilon_{\rm s} \dot{m}(t+t_0) \ast g(t) + \dot{m}(t),
\end{eqnarray}

\noindent
where $x_{\rm h}(t)$ and $x_{\rm s}(t)$ are the hard- and soft X-ray light curves, respectively, $\dot{m}$ is the mass accretion rate, $t_{\rm 0}$ is the time delay and $g(t)$ is the low-pass filter \citep[see][for further details]{Veledina18}. The variables $\varepsilon_{\rm h}$ and $\varepsilon_{\rm s}$ parameterise the contribution of the disc and synchrotron Comptonization components, and their signs indicate the correlation (plus) or anti-correlation (minus) with the accretion rate fluctuations. The optical light curve is given by 

\begin{eqnarray}
  o(t) &=&  - \dot{m}(t) + \varepsilon_{\rm ds} x_s(t) \ast r(t), 
\end{eqnarray}

\noindent
where the first term gives the synchrotron contribution and $r(t)$ is the disc response function whose contribution is parameterised by $\varepsilon_{\rm ds}$ \citep{Veledina11}. An implicit parameter of the model is the assumed shape of the accretion rate power spectrum, which  greatly affects the exact shape (and width) of the features in the CCF. We take a single zero-centered Lorentzian for each model. The resulting soft- and hard-band CCFs and are shown in Figure\,\ref{fig:model}.

We do not attempt to fit the 2019 March 2 and 2019 June 7 part 2  observations because the CCFs are at the noise level. However, the 2019 March 4 and 2019 June parts 1 and 4 CCFs show a nearly symmetric positive correlations at positive lags consistent with simple X-ray disc reprocessing. The soft- and hard-band CCFs from 2019 March 4, June 7 parts 1 and 4 are similar in shape indicating that one component dominates both the soft and hard X-rays. These CCFs can be reproduced if disc reprocessing dominates in the optical ($\varepsilon_{\rm ds}= 3$) and when Comptonization of either synchrotron or disc photons dominate in the X-rays. We find that the resulting CCFs (Fig.\,\textcolor{blue}{\ref{fig:model}a}) can be reproduced  with both $\varepsilon_{\rm h}=\varepsilon_{\rm s}\gg1$ and  $\varepsilon_{\rm h}=\varepsilon_{\rm s}\ll1$, hence the data do not allow us to distinguish between these components. We assume $\varepsilon_{\rm h}=\varepsilon_{\rm s}=0.01$.

We note that the 2019 June 7 part 3 data was observed between parts 1 and 4 in time, but shows a very  different soft- and hard- X-ray  CCF. It shows a positive correlations at negative lag in the soft-band, while in the hard X-ray band the correlation amplitudes are not significant. To reproduce the positive correlation at negative lags we consider an alternative scenario where the optical synchrotron emission is correlated with the hot flow X-ray emission, which translates to the formal description

\begin{equation}
  o(t) = \dot{m}(t) + \varepsilon_{\rm ds} x_{\rm s}(t) \ast r(t). 
\end{equation}

\noindent
This corresponds to the case when the fluctuations in the magnetic field (as response to the mass accretion rate variations) dominates the fluctuations in the number density and the emission in the optical band falls below the synchrotron self-absorption frequency \citep[see Fig.\,1b of][]{PV2009}. The resulting CCF is shown in  Fig.\,\textcolor{blue}{\ref{fig:model}e} and is obtained assuming the parameters $(\varepsilon_{\rm ds}, \varepsilon_{\rm s})$ to be (0,-0.8), i.e. assuming  contribution of the synchrotron alone to the optical band. The parameter $\varepsilon_{\rm h}$ is not constrained, as the hard-band CCF does not show any significant correlations. The parameter $t_0$ mildly affects the fit; we assume $t_{\rm 0}=4$.

Overall, we see a general trend in the  short-term variability. In the 2019 March and parts 1 and 4 of the June data, we find the optical flux is dominated by a reprocessing component (with possible minor contribution from the hot flow synchrotron emission), and the X-rays are also dominated by one component -- either synchrotron- or disc Comptonization. For the  2019 June 7 part 3 data we see an indication of a positive correlation at negative lags in the optical/soft-band X-ray CCF (see Fig.\,\textcolor{blue}{\ref{fig:ccf_energy}e}). The lack of a significant correlation in the optical/hard X-rays may be explained by the competing role of two anti-correlated spectral components in the X-rays, which we attribute to disc and synchrotron Comptonization. We find that the role of synchrotron emission in the optical band increases. 
This scenario can also explain the changing width of the ACFs. The 2019 March 4 data taken during the hard state have optical ACFs that are broader than the X-ray ACFs, consistent with X-ray reprocessing. The soft- and hard-band X-ray CCFs are similar implying that a single component dominates the X-rays. In contrast, the 2019 June part 3 data has a strong soft X-ray component with comparable optical and X-ray ACFs. This is indicative of the interplay between two anti-correlating components. The optical/soft- and optical/hard X-ray  CCFs are very different reflecting the comparable role of disc Comptonization and synchrotron Comptonization.

\begin{figure}
\includegraphics[width=0.50\textwidth]{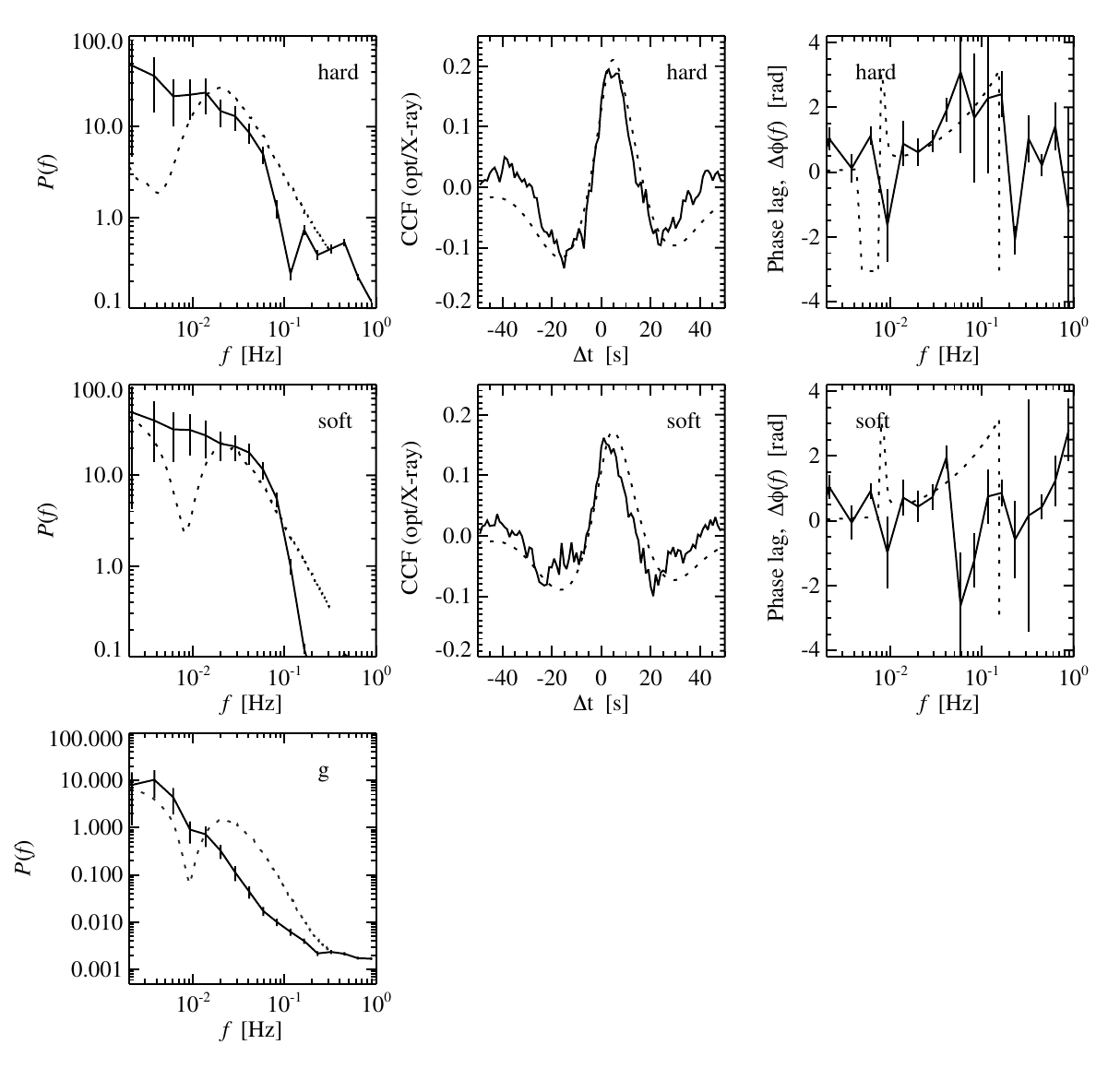}
\caption{We show the power spectrum, CCF and phase lags for the 2019 March 4 (b) data (solid line) with the corresponding hybrid hot inner flow model (dashed line). }
\label{fig:model_all} 
\end{figure}

\bsp	
\label{lastpage}
\end{document}